\documentclass[11pt]{article}

\usepackage[final]{acl}

\usepackage{times}
\usepackage{latexsym}
\usepackage{natbib}
\usepackage[T1]{fontenc}
\usepackage[utf8]{inputenc}

\usepackage{microtype}

\usepackage{inconsolata}

\usepackage{graphicx}

\usepackage{amsmath}
\usepackage{booktabs}
\usepackage{amssymb}
\usepackage{multirow}
\usepackage{tabularx}
\usepackage{array}
\usepackage{xcolor}  % in the preamble

\usepackage[most]{tcolorbox}
\usepackage{placeins}
\tcbuselibrary{breakable}

\newtcolorbox{hallucinationcase}[1][]{
  colback=gray!3,
  colframe=black!40,
  fonttitle=\bfseries,
  title=#1,
  breakable=false,
  boxrule=0.5pt,
  arc=2pt,
  left=6pt,
  right=6pt,
  top=6pt,
  bottom=6pt
}

\newcolumntype{M}{>{\centering\arraybackslash}X}

\title{HalluAudio: A Comprehensive Benchmark for Hallucination Detection in Large Audio-Language Models}

\author{
 \textbf{Feiyu Zhao\textsuperscript{1}\thanks{Equal contribution}},
 \textbf{Yiming Chen\textsuperscript{2}\footnotemark[1]},
 \textbf{Wenhuan Lu\textsuperscript{1}},
 \textbf{Daipeng Zhang\textsuperscript{1}},
 \\
 \textbf{Xianghu Yue\textsuperscript{1}\thanks{Corresponding author}},
 \textbf{Jianguo Wei\textsuperscript{1}}
\\
\\
 \textsuperscript{1}College of Intelligence and Computing, Tianjin University, China
\\
 \textsuperscript{2}ASUS Intelligent Cloud Services, Singapore
 \\
 \texttt{\textsuperscript{1}\{zhaofeiyu, wenhuan, zhangdaipeng, yuexianghu, jianguo\}@tju.edu.cn} \\
 \texttt{\textsuperscript{2}MattYM\_Chen@asus.com}
}

\begin{document}
\maketitle
\begin{abstract}
Large Audio-Language Models (LALMs) have recently achieved strong performance across various audio-centric tasks. 
However, hallucination, where models generate responses that are semantically incorrect or acoustically unsupported, remains largely underexplored in the audio domain. 
Existing hallucination benchmarks mainly focus on text or vision, while the few audio-oriented studies are limited in scale, modality coverage, and diagnostic depth.
We therefore introduce \textit{HalluAudio}, the first large-scale benchmark for evaluating hallucinations across speech, environmental sound, and music. 
\textit{HalluAudio} comprises over 5K human-verified QA pairs and spans diverse task types, including binary judgments, multi-choice reasoning, attribute verification, and open-ended QA. 
To systematically induce hallucinations, we design adversarial prompts and mixed-audio conditions. 
Beyond accuracy, our evaluation protocol measures hallucination rate, yes/no bias, error-type analysis, and refusal rate, enabling a fine-grained analysis of LALM failure modes. 
We benchmark a broad range of open-source and proprietary models, providing the first large-scale comparison across speech, sound, and music.
Our results reveal significant deficiencies in acoustic grounding, temporal reasoning, and music attribute understanding, underscoring the need for reliable and robust LALMs. \footnote{https://github.com/Feiyuzhao25/halluaudio}
\end{abstract}

% \ym{Use paragraph instead of textbf}

\section{Introduction}
Large Language Models (LLMs) have driven rapid progress in natural language processing, achieving strong performance across tasks such as reasoning, question answering, and multimodal understanding. Building on these advances, Large Audio-Language Models (LALMs) have emerged as a natural extension of LLMs to the audio domain. By leveraging large-scale corpora of speech, environmental sounds, and music, LALMs demonstrate impressive capabilities in speech recognition \cite{chu2023qwen, fang2025llama}, sound question answering \cite{xu2025qwen2, goel2025audio}, and music understanding \cite{ghosh2025music, coreteam2025mimoaudio}.
As LALMs are increasingly deployed in real-world applications, their accuracy and reliability have become critical. In particular, hallucinations remain a major concern, where models generate responses that are semantically incorrect or unsupported by the underlying audio.

% However, hallucination remains a persistent challenge for LLMs. 
A substantial body of work has demonstrated that generative models often produce outputs that are fluent yet factually unsupported in both text-only~\citep{li2023halueval,lin2022truthfulqa} and vision–language settings~\citep{guan2024hallusionbench,wu2024autohallusion}.
In contrast, hallucination in audio-centric models remains largely underexplored.
Most existing benchmarks focus on text or vision, while the few emerging studies in the audio domain are limited in scale, modality coverage, and task diversity~\citep{chengaha}. Current evaluations typically rely on small binary classification tasks and rarely probe critical failure modes such as response bias, refusal behavior, or multi-turn inconsistency. As a result, the field of LALMs still lacks a dedicated large-scale benchmark for systematically characterizing hallucination across speech, environmental sound, and music tasks.

To bridge this gap, we introduce \textit{HalluAudio}, the first large-scale human-verified benchmark suite specifically designed to evaluate hallucination in LALMs.
\textit{HalluAudio} spans three major audio domains, speech, environmental sounds, and music, and supports diverse task formats, including classification, question answering, and open-ended generation.
To reliably elicit and measure hallucinations, we incorporate adversarial prompts, mixed audio conditions.
To ensure the reliability of our evaluation suite, we also perform manual inspection on the dataset. We curate \textit{HalluAudio} through a five-step pipeline including data collection, task construction, adversarial augmentation, automated filtering, and multi-round human verification, resulting in a large-scale benchmark spanning three audio domains, dozens of task types, and over 5K carefully validated QA pairs.

We evaluate a range of state-of-the-art open source and proprietary LALMs on HalluAudio.
Comprehensive experiments show that hallucination remains a systematic issue in current LALMs, even for tasks with a clear audio answer. We observe consistent Yes/No biases, non-trivial false refusal behaviors, and domain-dependent failure patterns across speech, sound, and music. More importantly, strong performance on standard audio benchmarks does not necessarily imply robustness against hallucination, highlighting a gap between capability evaluation and reliability assessment. These findings underscore the need for targeted hallucination diagnostics and validate \textit{HalluAudio} as an effective tool for analyzing fine-grained failure modes beyond aggregate accuracy.

Our main contributions are as follows:

\begin{itemize}
\item \textbf{A large-scale human-verified benchmark for audio hallucination.}
We introduce \textit{HalluAudio}, the first large-scale human-verified audio hallucination benchmark spanning speech, environmental sounds, and music, with thousands of QA pairs per domain covering both audio understanding and audio-grounded generation scenarios.

\item \textbf{A diverse suite of hallucination-inducing task designs.}
\textit{HalluAudio} includes binary and multi-class classification, audio QA, attribute verification, comparative reasoning, and open-ended generation. We construct adversarial prompts, mixed-audio inputs, and positives/negatives to systematically trigger and measure hallucinations.

\item \textbf{A multi-dimensional empirical analysis of hallucination in LALMs.}
We evaluate a broad set of open-source and proprietary LALMs and report detailed hallucination patterns across tasks and modalities. Our analysis incorporates accuracy, hallucination rate, Yes/No bias, error-type breakdown, and refusal rate, revealing critical reliability gaps in current LALMs.

\end{itemize}

\section{Related Works}

\paragraph{Large-Audio Language Models} 
Inspired by the success of LLMs, recent work integrates audio representations with LLMs, giving rise to LALMs that process speech, environmental sounds, and music to generate textual responses and reasoning outputs. Early efforts such as AudioLM~\citep{borsos2023audiolm} demonstrated that discretized audio tokens can be effectively modeled with language modeling techniques, enabling coherent continuation of speech and music. Subsequent models, including SALMONN~\citep{tangsalmonn}, Qwen2-Audio~\citep{chu2024qwen2}, and Audio Flamingo~\citep{kong2024audio}, further align audio encoders with LLMs to support tasks such as ASR, audio question answering, captioning, and music understanding. 
More recent studies extend LALMs to finer-grained music understanding~\citep{huang2022mulan} and improved instruction following~\citep{frieske2024hallucinations}.
Overall, LALMs are rapidly evolving from transcription-focused systems to general audio reasoning models. 
Although several benchmarks have been proposed to evaluate various capabilities of LALMs~\citep{yang2024air,chen-etal-2024-beyond-single,chen2026voicebench,chen2026loasr}, their reliability, particularly with respect to hallucination, remains underexplored.

\begin{figure*}[htbp]
\centering
\includegraphics[width=1\textwidth]{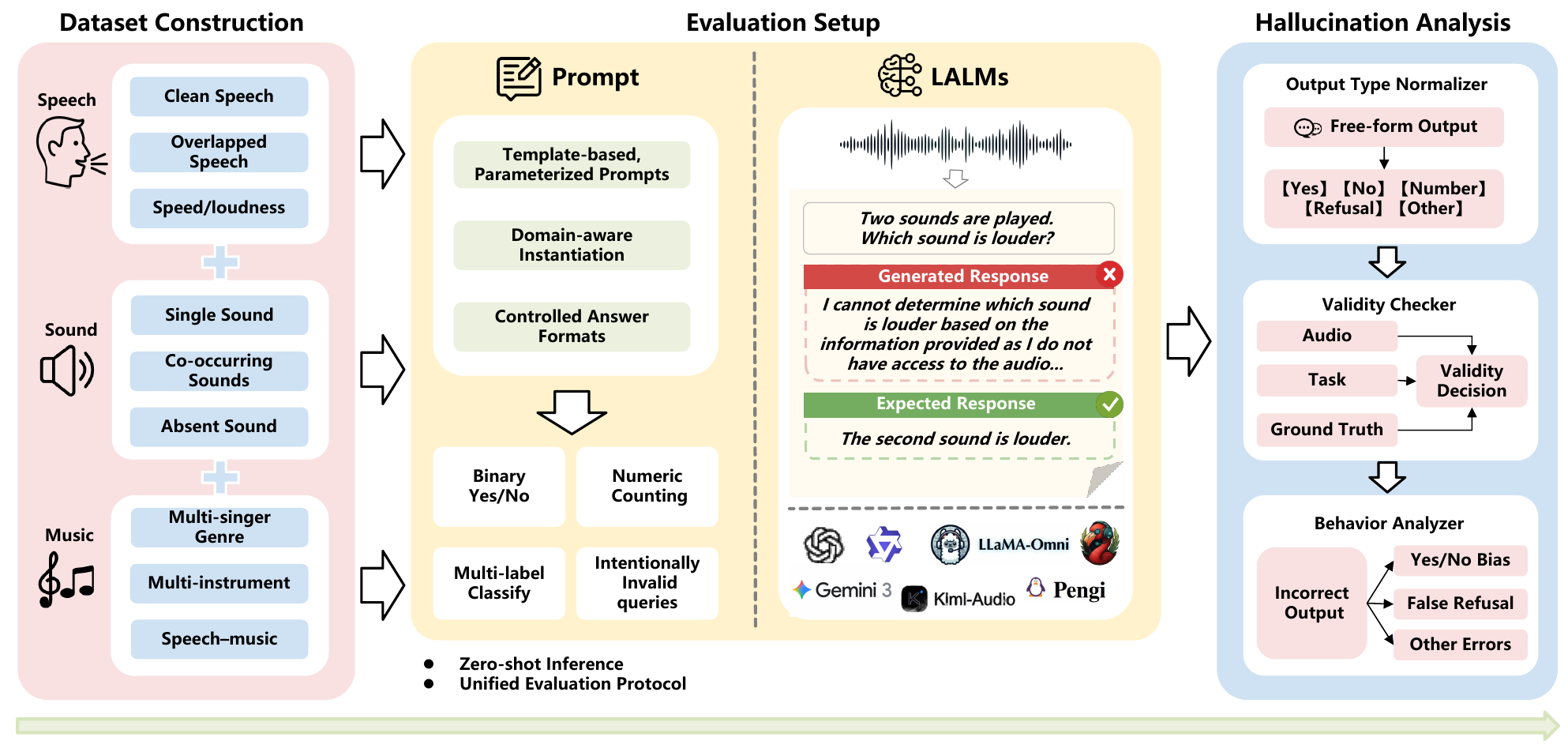}
\caption{Overview of the \textit{HalluAudio} framework.
\textit{HalluAudio} combines controlled multi-domain audio construction, unified prompting, and structured output validation to systematically analyze hallucination in LALMs.}
\label{framework}
\end{figure*}

\paragraph{Hallucinations in Audio Tasks} 
Hallucination has been extensively studied in text \cite{bang2025hallulens} and vision \cite{cao2024visdiahalbench} domains, revealing systematic failures in grounding, object hallucination, and cross-modal consistency.
In contrast, hallucinations in the audio domain remain largely underexplored. 
\citet{frieske2024hallucinations} provides an early analysis of hallucination in ASR, showing that metrics such as WER fail to detect fluent but semantically irrelevant outputs and exposing vulnerabilities to misleading acoustic cues. 
AHa-Bench~\citep{chengaha} extends this line of inquiry to LALMs through a small-scale binary QA benchmark. 
While informative, existing benchmarks are limited in dataset scale, task diversity, and diagnostic depth, leaving critical failure modes such as response bias and refusal behavior. 
The field still lacks a unified taxonomy, controlled contrastive audio pairs, multi-format evaluation tasks, and large-scale human-annotated assessments. HalluAudio addresses these gaps by introducing the first large-scale, multi-domain, and multi-dimensional benchmark for hallucination in LALMs.
% \ym{Add definition of hallucination. The importance of investigating hallucination. Add discussion on audio hallucination papers, e.g., 1. Reducing Object Hallucination in Large Audio-Language Models via Audio-Aware Decoding, 2. Can Large Audio-Language Models Truly Hear? Tackling Hallucinations with Multi-Task Assessment and Stepwise Audio Reasoning 3. Teaching audio-aware large language models what does not hear: Mitigating hallucinations through synthesized negative QA pairs}

% These include TruthfulQA \cite{lin2022truthfulqa} for probing adversarial factual errors, HaluEval \cite{li2023halueval} for detecting fabricated dialogue content, C-FAITH \cite{zhang2025c} for fine-grained Chinese hallucination assessment, HaluLens \cite{bang2025hallulens} for dynamically generating external hallucination tasks, and DefAn \cite{rahman2025defan} for evaluating models’ ability to avoid unwarranted definitive claims. In the vision-language domain, benchmark suites such as HallusionBench \cite{guan2024hallusionbench}, AutoHallusion \cite{wu2024autohallusion}, R-Bench \cite{wu2024evaluating}, and NOPE \cite{lovenia2024negative} analyze object hallucination, cross-modal inconsistency, and visually grounded reasoning failures in Large Vision-Language Models (LVLMs). Together, these efforts demonstrate that hallucination is pervasive across both language and multimodal models, driving the continued need for systematic diagnostic evaluation.

% \section{HalluVoice: A Benchmark for Evaluating Hallucinations in LALMs}
\section{HalluAudio Benchmark}

\subsection{Overview of \textit{HalluAudio}}

The overall evaluation pipeline of HalluAudio is illustrated in Figure 1, which follows a modular, end-to-end process designed to systematically elicit, measure, and analyze hallucinations in LALMs. Specifically, given curated audio inputs from three domains: speech, sound, and music, we construct diverse task instances via template-based, parameterized prompts with domain-aware instantiation and controlled answer formats. These audio–prompt pairs are evaluated under a unified zero-shot protocol across a suite of LALMs, producing textual outputs in heterogeneous forms, including binary decisions, numeric counts, and free-form responses. Finally, model outputs are then normalized into structured types and validated against task definitions and ground-truth audio evidence through an automated evaluation engine. Based on the validated outcomes, we conduct fine-grained hallucination analysis across domains and tasks, covering Yes/No bias, false refusals, and other error patterns.

In this work, we define audio hallucination operationally as a model-generated claim that is not supported by the acoustic evidence in the input. This includes three representative cases: (1) fabrication, where the model asserts the presence of non-existent audio events; (2) evidence contradiction, where the response conflicts with deterministic acoustic structure; and (3) unjustified affirmative bias, where the model produces positive responses despite insufficient or absent evidence. This definition explicitly distinguishes hallucination from general capability errors, such as failures due to reasoning complexity or ambiguous inputs.

Based on this definition, HalluAudio organizes evaluation into three hallucination-sensitive dimensions aligned with task design: (1) \emph{Structural and temporal hallucinations}, evaluated by \textbf{Temporal Comparison} tasks, where models make incorrect claims about order, timing, or quantitative acoustic properties; (2) \emph{Perceptual hallucinations}, captured by \textbf{Recognition} tasks, involving incorrect assertions about the presence or attributes of sounds or musical elements; and (3) \emph{Semantic hallucinations}, assessed through \textbf{Consistency} tasks, where model responses become internally inconsistent or unsupported under controlled input perturbations. Our objective is to provide a systematic, multi-domain evaluation protocol for analyzing hallucination behaviors in modern LALMs.

\begin{table}[htbp]
\centering
\small
\setlength{\tabcolsep}{6pt}
\begin{tabular}{l c}
\toprule
\textbf{Domain} & \textbf{Dataset}\\
\midrule
Speech &
Common Voice \cite{ardila2020common} \\
\addlinespace

Sound &
FSD50K \cite{fonseca2021fsd50k} \\
\addlinespace

\multirow{3}{*}{Music} &
GTZAN \cite{sturm2012analysis}\\
&
Mridangam Strokes \cite{turian2022hear} \\
&
Mridangam Tonics \cite{turian2022hear} \\
\bottomrule
\end{tabular}
\caption{Audio sources used in HalluAudio.}
\label{tab:dataset_sources}
\end{table}

\begin{figure*}[htbp]
\centering
\includegraphics[width=1\textwidth]{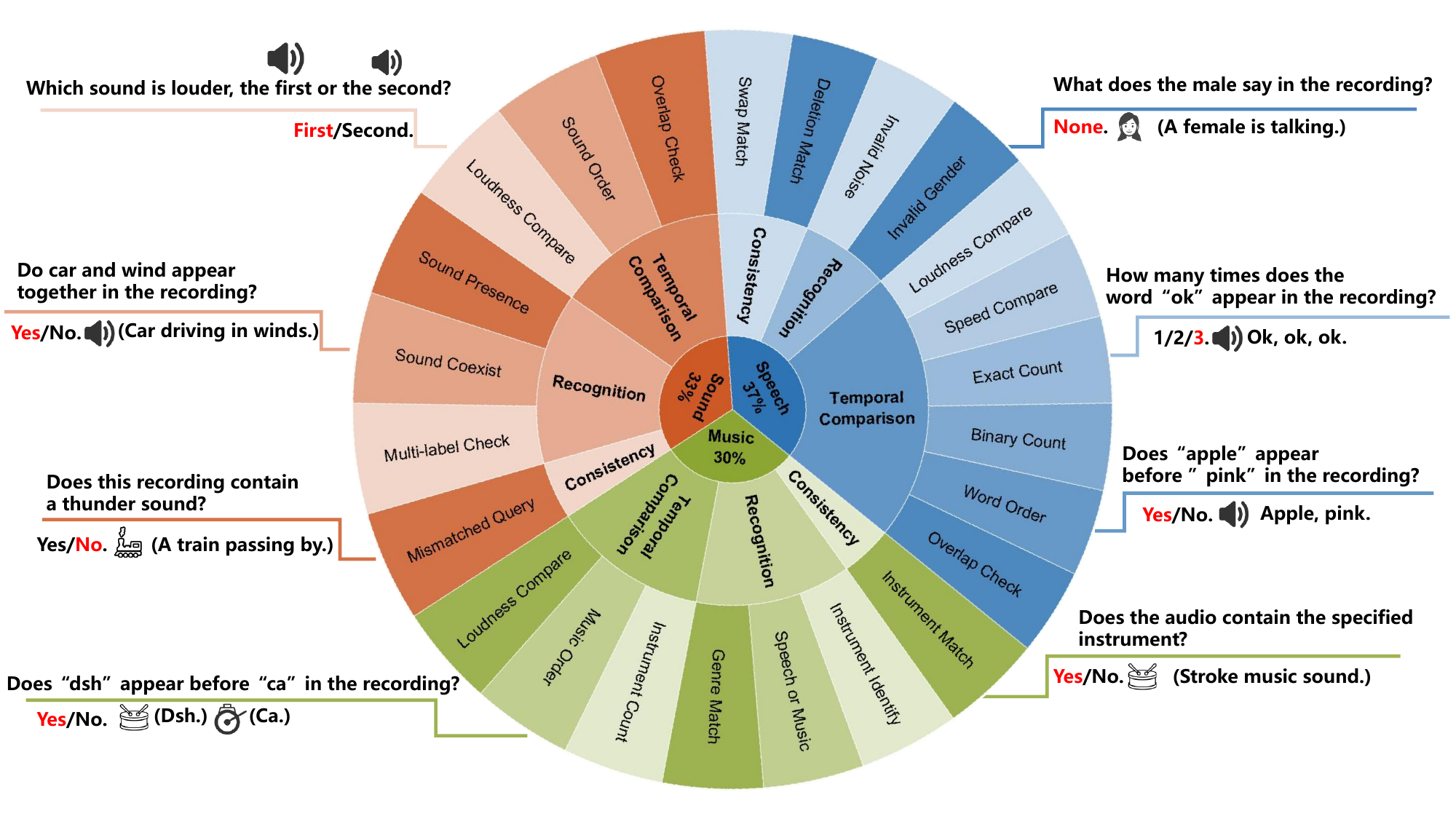}
\caption{Detailed task composition of the HalluAudio dataset across speech, sound, and music domains.} 
\label{dataset}
\end{figure*}

\subsection{Dataset Construction}

We construct HalluAudio through a controlled, reproducible pipeline designed to elicit and diagnose hallucination behaviors in LALMs.

\paragraph{Step 1. Audio selection}  
We curate audio clips from speech, environmental sound, and music corpora with reliable annotations. Clips are selected to cover diverse acoustic conditions, including multi-speaker speech, event-rich sound, and structured musical segments. Audio sources are summarized in Table~\ref{tab:dataset_sources}.

\paragraph{Step 2. Template-based prompt generation}  
For each task, we design parameterized prompt templates with slot variables. Slots are instantiated using clip annotations for valid queries or deliberately mismatched attributes for invalid queries, producing large-scale audio–question pairs. 

\paragraph{Step 3. Contrastive and adversarial construction}  
We generate paired instances by minimally modifying prompts or audio attributes, ensuring controlled positive/negative contrasts that isolate hallucination triggers. Audio attributes cover temporal order, event presence, loudness, counting, and musical structure, while prompt modifications include adversarial negatives, invalid queries, and attribute-level perturbations.

\paragraph{Step 4. Validation and quality control}  
We first generate a large pool of candidate instances using automated scripts. Each instance then undergoes three rounds of human verification involving two independent annotators and one senior reviewer. Annotators are instructed to check: (1) alignment between audio content and the question, (2) correctness of the ground-truth answer, and (3) whether the instance satisfies the intended hallucination-triggering condition. Disagreements are resolved through majority voting and adjudication by the senior reviewer. QA pairs with persistent ambiguity or inconsistent interpretations are revised or discarded to ensure high dataset reliability. The annotators are three PhD-level researchers specializing in speech and audio processing, ensuring domain expertise during verification.

\paragraph{Step 5. Packaging and balancing} 
In final step, we further balance the dataset across domains, task types, and hallucination categories to avoid distributional bias and ensure fair evaluation of different hallucination behaviors.

\begin{table*}[t]
\centering
\small
\setlength{\tabcolsep}{4pt}
\renewcommand{\arraystretch}{1.2}
\begin{tabular}{l l r p{9.5cm} l}
\toprule
Domain & Category & Count & Prompt Template & \textbf{\#O-QA} \\
\midrule
\multirow{10}{*}{Speech}
& Overlap Check      & 189 & Do the two speakers' voices overlap in the recording? & $\times$ \\
& Word Order         & 245 & Does the word ``[AAA]'' appear before the word ``[BBB]'' in the recording? & $\times$ \\
& Binary Count       & 156 & Does the speaker say the word ``[AAA]'' [XXX] times in the recording? & $\times$ \\
& Exact Count        & 178 & How many times does the word ``[AAA]'' appear in the recording? & \checkmark \\
& Invalid Gender     & 172 & What does the (fe)male say in the recording? & \checkmark \\
& Invalid Noise      & 192 & What does the speaker say in the recording? & \checkmark \\
& Deletion Match     & 303 & Does the speech recording match the transcription: ``[AAA]''?  & $\times$ \\
& Swap Match         & 310 & Does the speech recording match the transcription: ``[AAA]''?  & $\times$ \\
& Speed Compare      & 230 & Which instance of ``[AAA]'' was spoken faster, the first or the second? & \checkmark \\
& Loudness Compare   & 225 & Which instance of ``[AAA]'' was spoken louder, the first or the second? & \checkmark \\
\midrule
\multirow{7}{*}{Sound}
& Overlap Check      & 254 & Do ``[LABEL1]'' and ``[LABEL2]'' overlap in the recording? & $\times$ \\
& Sound Order         & 300 & Does ``[LABEL1]'' appear before ``[LABEL2]'' in the recording? & $\times$ \\
& Sound Presence     & 260 & Does the recording contain a ``[LABEL]'' sound? & $\times$ \\
& Sound Coexist      & 300 & Do ``[LABEL1]'' and ``[LABEL2]'' appear together in the recording? & $\times$ \\
& Mismatched Query    & 257 & Does this recording contain a ``[RANDOM\_LABEL]'' sound? & $\times$ \\
& Multi-label Check  & 287 & Are there multiple sound types in this recording? & $\times$ \\
& Loudness Compare   & 300 & Which sound is louder, the first or the second?  & \checkmark \\

\midrule
\multirow{7}{*}{Music}
& Genre Match        & 291 & Is this audio clip [LABEL] music?  & $\times$ \\
& Instrument Match   & 258 & Does the audio contain the specified instrument or stroke label?  & $\times$ \\
& Speech or Music   & 128   & Is this audio clip speech or music? & \checkmark \\
& Instrument Identify      & 233 & Is the stroke or tonic type [LABEL]?  & $\times$ \\
& Loudness Compare   & 252 & Which sound is louder, the first or the second?  & \checkmark \\
& Music Order         & 297 & Does ``[LABEL1]'' appear before ``[LABEL2]'' in the recording?  & $\times$ \\
& Instrument Count & 300 & How many sound types in this recording?  & \checkmark \\
\midrule
Total & - & 5720 & - & - \\
\bottomrule
\end{tabular}
\caption{Detailed task composition of the HalluAudio dataset across speech, environmental sound, and music domains. \textbf{\#O-QA}: open-ended QA pairs.}
\label{tab:dataset_details}
\end{table*}

\subsection{Dataset Statistics}

% \textbf{Detailed Statistical Analysis of HalluVoice.} 
As illustrated in Figure \ref{dataset}, HalluAudio covers three audio domains with a diverse set of task categories and balanced sample distributions. The speech domain contains the largest variety of fine-grained tasks, including temporal reasoning, transcription consistency checks, comparative judgments, and invalid or underspecified queries, reflecting the complexity of speech-based hallucination behaviors. Environmental sound tasks emphasize sound presence, co-occurrence, and adversarial negative queries, while music tasks focus on instrument and genre identification, comparative reasoning, and cross-domain invalid prompts.

HalluAudio explicitly incorporates contrastive and adversarial constructions to enable targeted hallucination diagnosis. Contrastive tasks account for 2,662 out of 5,720 QA pairs, while explicitly adversarial or invalid queries account for 621 QA pairs. In total, 57.4\% of the dataset is designed to probe hallucination through controlled perturbations or absence of evidence, distinguishing HalluAudio from conventional audio QA datasets that primarily contain valid and answerable queries.

\paragraph{Task Composition and Statistics}

Table~\ref{tab:dataset_details} presents a detailed breakdown of the HalluAudio dataset. 
The benchmark spans three domains, speech, environmental sound, and music, and covers a diverse set of task categories designed to probe temporal reasoning, counting, matching, comparison, and invalid or underspecified queries. 
Each task is instantiated using parameterized prompt templates with slot variables, enabling large-scale construction while maintaining precise alignment between audio evidence and ground-truth answers. 
Invalid query categories intentionally lack sufficient audio support and are used to diagnose hallucination behaviors such as overconfident guessing or inappropriate refusals.

\paragraph{Comparison with Other Benchmarks.}

\begin{table}[htbp]
\centering
% \small
% \setlength{\tabcolsep}{2pt}
\resizebox{\columnwidth}{!}{
\begin{tabular}{lccccc}
\toprule
\textbf{Benchmarks} 
& \textbf{\#Ds-D}
& \textbf{\#H-E}
& \textbf{\#O-QA} 
& \textbf{Scale} \\
\midrule

USMQ~\cite{kuan2024understanding}
& $\times$ 
& $\times$ 
& $\times$ 
& {$\sim$30K} \\

Match~\cite{kuan2025can}
& $\times$ 
& $\times$ 
& $\times$ 
& {$>$15K} \\

Avhbench~\cite{sungavhbench}
& $\times$ 
& $\times$ 
& $\times$ 
& $\sim$5K \\

AHa-Bench~\cite{chengaha}
& $\times$ 
& \checkmark
& $\times$ 
& $\sim$1K \\

\textbf{HalluAudio (ours)} 
& \checkmark
& \checkmark
& \checkmark 
& \textbf{$>$5K} \\
\bottomrule
\end{tabular}
}
\caption{Comparison of different hallucination benchmarks for LALMs. \textbf{\#Ds-D}: Domain-specific Design. \textbf{\#H-E}: Number of manually verified QA pairs. \textbf{\#O-QA}: Open-ended QA.}
\label{dataset_comparison}
\end{table}

\begin{table*}[t]
\centering
\small
\setlength{\tabcolsep}{1.5pt}
\renewcommand{\arraystretch}{1.25}
\begin{tabularx}{\textwidth}{l *{10}{M} M}
\toprule
\multirow{2}{*}{\textbf{Model}} &
\multicolumn{6}{c}{\textbf{Temporal Comparison}} &
\multicolumn{2}{c}{\textbf{Recognition}} &
\multicolumn{2}{c}{\textbf{Consistency}} &
\multirow{2}{*}{\textbf{Average}} \\
\cmidrule(lr){2-7}
\cmidrule(lr){8-9}
\cmidrule(lr){10-11}
&
\textbf{overlap} &
\textbf{order} &
\textbf{speed} &
\textbf{loudness} &
\textbf{exact} &
\textbf{binary} &
\textbf{noise} &
\textbf{gender} &
\textbf{match\_s} &
\textbf{match\_d} &
\\
\midrule

Qwen-Audio           & 55.78 & 46.81 & \textbf{51.50} & 48.97 & 44.58 & 43.57 & 1.86 & 0.13 & 14.94 & 13.41 & 32.16 \\
Qwen2-Audio          & 50.00 & 51.46 & 3.35  & \underline{0.00} & 6.57  & 47.31 & \underline{0.00} & 18.97 & 58.71 & 50.83 & 28.72 \\
Llama-Omni           & 41.26 & 38.11 & \underline{0.00} & 0.04  & 3.92  & 37.28 & 0.11 & \underline{0.00} & 30.96 & 30.96 & 18.26 \\
Llama-Omni2          & 50.38 & 55.38 & 21.87 & 20.02 & 48.72 & 56.39 & \textbf{99.93} & 12.94 & 17.36 & 26.36 & 40.94 \\
Kimi-Audio           & 52.27 & 51.73 & 29.78 & 46.49 & 48.87 & 58.53 & 8.40 & \textbf{82.57} & 15.18 & 13.47 & 40.73 \\
Phi-4-Multimodal     & 54.07 & 57.40 & 19.40 & 29.35 & 69.15 & 59.32 & \underline{0.00} & 0.85  & 39.75 & 25.42 & 35.47 \\
Pengi                & 50.00 & \underline{34.10} & \underline{0.00} & \underline{0.00} & 20.58 & 39.26 & \underline{0.00} & \underline{0.00} & \underline{11.74} & 11.51 & 16.72 \\
MiMo-Audio           & 96.30 & \textbf{79.59} & 24.78 & 47.56 & 57.87 & 55.13 & 2.62 & 0.58  & \textbf{91.22} & \textbf{72.73} & \textbf{52.84} \\
Step-Audio-2         & \textbf{99.47} & 61.22 & 50.00 & \textbf{50.22} & 48.31 & 51.92 & 1.56 & 3.49  & 14.15 & \underline{2.53} & 38.29 \\
\midrule
GPT-4o-Audio         & 57.67 & 79.18 & 4.35  & 13.33 & \textbf{80.34} & \textbf{65.38} & 6.04 & 3.51  & 75.12 & 70.20 & 45.51 \\
Gemini-2.5-Flash     & \underline{9.84} & 38.59 & 1.74  & \underline{0.00} & \underline{2.26} & \underline{35.76} & 2.14 & 7.56  & 56.57 & 11.51 & \underline{16.60} \\

\bottomrule
\end{tabularx}
\caption{Classification accuracy (\%) on HalluAudio in speech domain.  \textbf{match\_s}: swap match. \textbf{match\_d}: deletion match.}
\label{speech_accuracy}
\end{table*}

Table~\ref{dataset_comparison} summarizes representative hallucination benchmarks for LALMs. Existing benchmarks such as USMQ \cite{kuan2024understanding}, Match \cite{kuan2025can}, and Avhbench \cite{sungavhbench} are of moderate scale and lack domain-specific task design, manual verification, or open-ended QA. AHa-Bench \cite{chengaha} provides manually verified QA pairs but is limited in scale and restricted to binary hallucination detection. In contrast, \textit{HalluAudio} adopts domain-specific task formulations across speech, environmental sound, and music, covering temporal, perceptual, and structural reasoning, and supports both binary and open-ended QA at a larger scale ($>$5K QA pairs), enabling more fine-grained analysis of hallucination behaviors across domains.

% \ym{Add a table comparing existing benchmarks with our benchmark to highlight what is new in our benchmark.}

\section{Evaluation}

% We evaluate LALMs under controlled settings, as detailed in the following subsections.

\subsection{Examined Models}

% We evaluate HalluAudio on a set of LALMs selected according to domain-specific capabilities, as existing LALMs are typically optimized for particular audio modalities rather than providing uniform coverage across domains.
We evaluate the performance of 12 LALMs, including 2 proprietary models: GPT-4o-Audio \cite{hurst2024gpt} and Gemini-2.5-Flash \cite{comanici2025gemini}, as well as 10 representative open-source
models: Qwen-Audio~\cite{chu2023qwen}, Qwen2-Audio~\cite{chu2024qwen2}, Qwen2.5-Omni~\cite{xu2025qwen2},
Llama-Omni~\cite{grattafiori2024llama},
Llama-Omni2~\cite{fang2025llama}, Kimi-Audio~\cite{ding2025kimi}, Phi-4-Multimodal~\cite{abouelenin2025phi}, Audio Flamingo 3~\cite{goel2025audio}, Music-Flamingo~\cite{ghosh2025music}, Pengi~\cite{deshmukh2023pengi}, MiMo-Audio~\cite{coreteam2025mimoaudio}, Step-Audio-2~\cite{wu2025step}. All models are evaluated with three independent runs, and reported results are averaged to ensure statistical stability.

\begin{table*}[t]
\centering
\small
\setlength{\tabcolsep}{1.5pt}
\renewcommand{\arraystretch}{1.25}
\begin{tabularx}{\textwidth}{l *{7}{M} M}
\toprule
\multirow{2}{*}{\textbf{Model}} &
\multicolumn{3}{c}{\textbf{Temporal Comparison}} &
\multicolumn{3}{c}{\textbf{Recognition}} &
\multicolumn{1}{c}{\textbf{Consistency}} & 
\multirow{2}{*}{\textbf{Average}}\\
\cmidrule(lr){2-4}
\cmidrule(lr){5-7}
\cmidrule(lr){8-8}
&
\textbf{overlap} &
\textbf{order} &
\textbf{loudness} &
\textbf{multi\_label} &
\textbf{presence} &
\textbf{coexist} &
\textbf{mismatch} & \\
\midrule 
Qwen2.5-Omni & 
87.97 & 88.89 & 92.77 & \textbf{75.35} & 87.97 & 61.68 & \textbf{98.17} & \textbf{84.69}\\
Audio Flamingo-3 &
63.53 & \textbf{98.70} & 95.25 & 13.27 & 26.95 & 72.13 & 83.67 & 64.79\\
Pengi &
\underline{23.90} & \underline{23.90} & \underline{8.67} & 15.47 & 34.34 & \underline{20.50} & 46.07 & \underline{24.69}\\
Kimi-Audio &
28.90 & 39.30 & 45.50 & \underline{11.80} & 93.61 & 45.80 & \underline{6.40} & 38.76\\
Qwen2-Audio &
64.63 & 94.47 & \textbf{95.78} & 48.23 & 97.60 & 59.41 & 11.43 & 67.36\\
MiMo-Audio &
\textbf{94.88} & 91.00 & 93.00 & 42.86 & 71.54 & 89.00 & 95.33 & 82.52 \\
Step-Audio-2 &
72.44 & 92.67 & 57.67 & 26.83 & \textbf{99.23} & \textbf{97.00} & 62.65 & 72.64\\
\midrule
GPT-4o-Audio &
66.53 & 85.52 & 72.67 & 56.10 & 95.35 & 81.00 & 39.84 & 71.00\\
Gemini-2.5-Flash &
25.53 & 27.96 & 40.77 & 51.45 & \underline{10.12} & 41.37 & 78.24 & 39.35 \\

\bottomrule
\end{tabularx}
\caption{Classification accuracy (\%) on HalluAudio in sound domain.}
\label{audio_accuracy}
\end{table*}

\begin{table*}[t]
\centering
\small
\setlength{\tabcolsep}{1.5pt}
\renewcommand{\arraystretch}{1.25}
\begin{tabularx}{\textwidth}{l *{9}{M} M}
\toprule
\multirow{2}{*}{\textbf{Model}} &
\multicolumn{4}{c}{\textbf{Temporal Comparison}} &
\multicolumn{3}{c}{\textbf{Recognition}} &
\multicolumn{2}{c}{\textbf{Consistency}} & 
\multirow{2}{*}{\textbf{Average}} \\
\cmidrule(lr){2-5}
\cmidrule(lr){6-8}
\cmidrule(lr){9-10}
&
\textbf{order} &
\textbf{count\_s} &
\textbf{count\_t} &
\textbf{loudness} &
\textbf{instru\_id} &
\textbf{genre} &
\textbf{s\_or\_m} &
\textbf{match\_s} &
\textbf{match\_t} & \\
\midrule
Qwen2.5-Omni &
49.38 & 25.80 & 14.84 & 95.40 & 43.92 & 62.16 & \textbf{100.00} & 34.07 & 33.07 & 50.96 \\
Audio Flamingo3 &
50.00 & 49.17 & 48.53 & \textbf{100.00} & 70.55 & 50.00 & 50.00 & 44.07 & 60.57 & 58.10 \\
Pengi &
\underline{0.00} & 51.37 & 29.23 & \underline{0.00} & 32.55 & 63.50 & \underline{0.00} & \underline{28.83} & 49.90 & \underline{28.38} \\
Kimi-Audio &
\textbf{100.00} & 29.53 & 20.17 & \textbf{100.00} & 32.90 & 50.45 & \underline{0.00} & 33.77 & 41.60 & 45.38 \\
Qwen2-Audio &
\underline{0.00} & \underline{4.20} & \underline{5.23} & \underline{0.00} & 53.30 & 51.85 & 50.00 & 51.07 & \underline{32.37} & 27.56 \\
Music-Flamingo &
50.00 & 28.11 & 20.42 & \textbf{100.00} & 53.80 & 50.00 & \textbf{100.00} & 30.80 & \textbf{73.60} & 56.30\\
MiMo-Audio &
\textbf{100.00} & \textbf{69.33} & \textbf{67.33} & \textbf{100.00} & 67.81 & 77.66 & \textbf{100.00} & \textbf{51.96} & 48.04 & \textbf{75.79}\\
Step-Audio-2 &
\textbf{100.00} & 38.00 & 36.67 & \underline{0.00} & \textbf{77.68} & \textbf{78.69} & 92.97 & 48.60 & 50.28 & 58.10\\
\midrule
GPT-4o-Audio &
99.66 & 21.33 & 14.67 & 75.40 & 67.81 & \textbf{79.86} & 99.22 & 28.49 & 36.87 & 58.15\\
Gemini-2.5-Flash &
13.45 & 12.70 & 11.02 & 8.64 & \underline{20.91} & \underline{45.87} & 87.50 & 40.67 & 42.86 & 31.51\\

\bottomrule
\end{tabularx}
\caption{Classification accuracy (\%) on HalluAudio in music domain. \textbf{count\_s}: instrument count in stroke. \textbf{count\_t}: instrument count in tonic. \textbf{instru\_id}: instrument identify. \textbf{s\_or\_m}: speech or music. \textbf{match\_s}: instrument match in stroke. \textbf{match\_t}: instrument match in tonic.}
\label{music_accuracy}
\end{table*}

% Table \ref{tab: LALMs} lists the relevant model selections under different domains \cite{chu2023qwen,chu2024qwen2,xu2025qwen2,grattafiori2024llama,fang2025llama,ding2025kimi,abouelenin2025phi,goel2025audio,ghosh2025music,deshmukh2023pengi,wu2025step,coreteam2025mimoaudio}. 
% All models are evaluated in a zero-shot setting with default inference parameters.

\begin{figure*}[htbp]
\centering
\setlength{\tabcolsep}{2pt}

\begin{tabular}{ccc}
\includegraphics[width=0.32\textwidth]{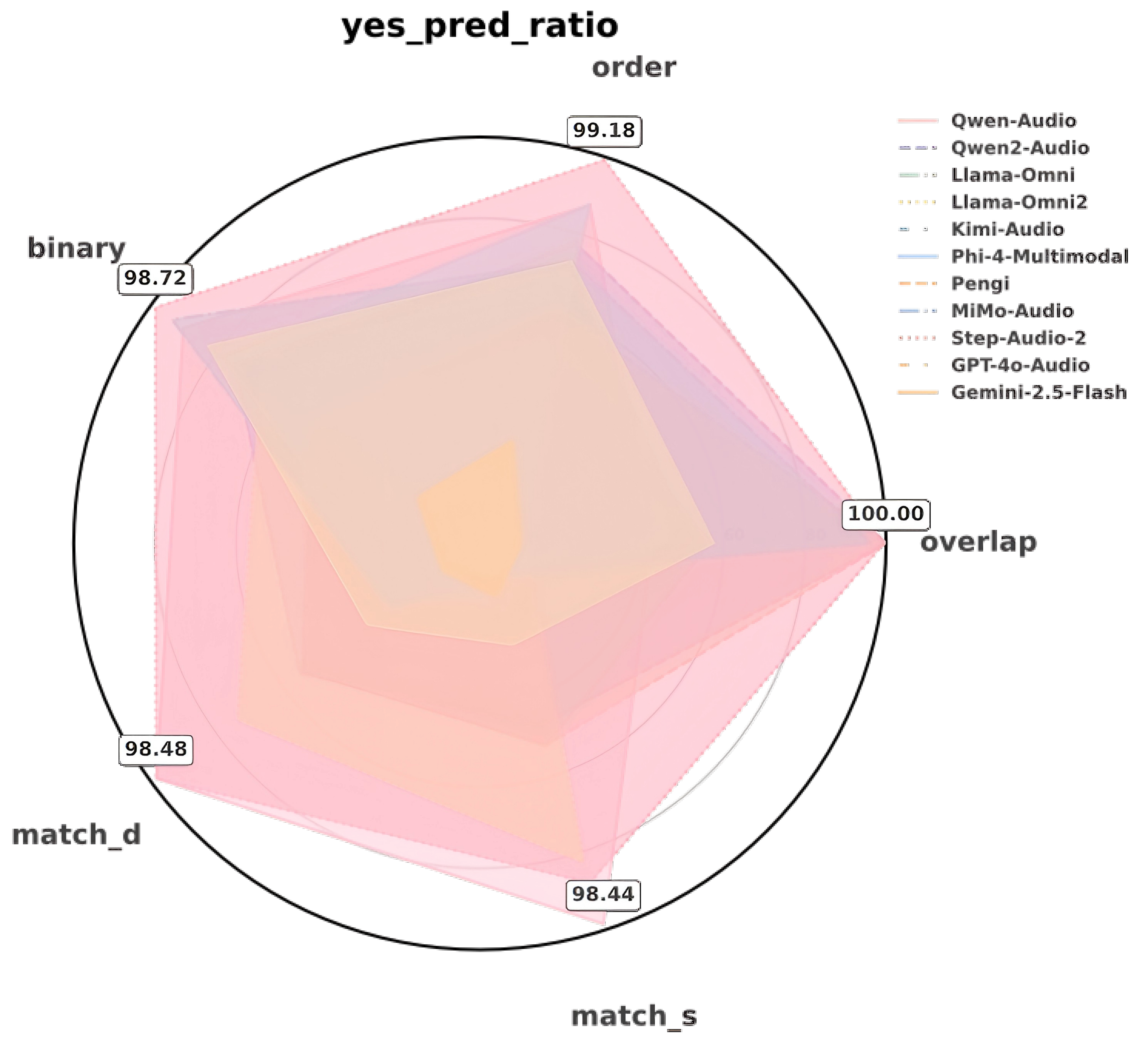} &
\includegraphics[width=0.32\textwidth]{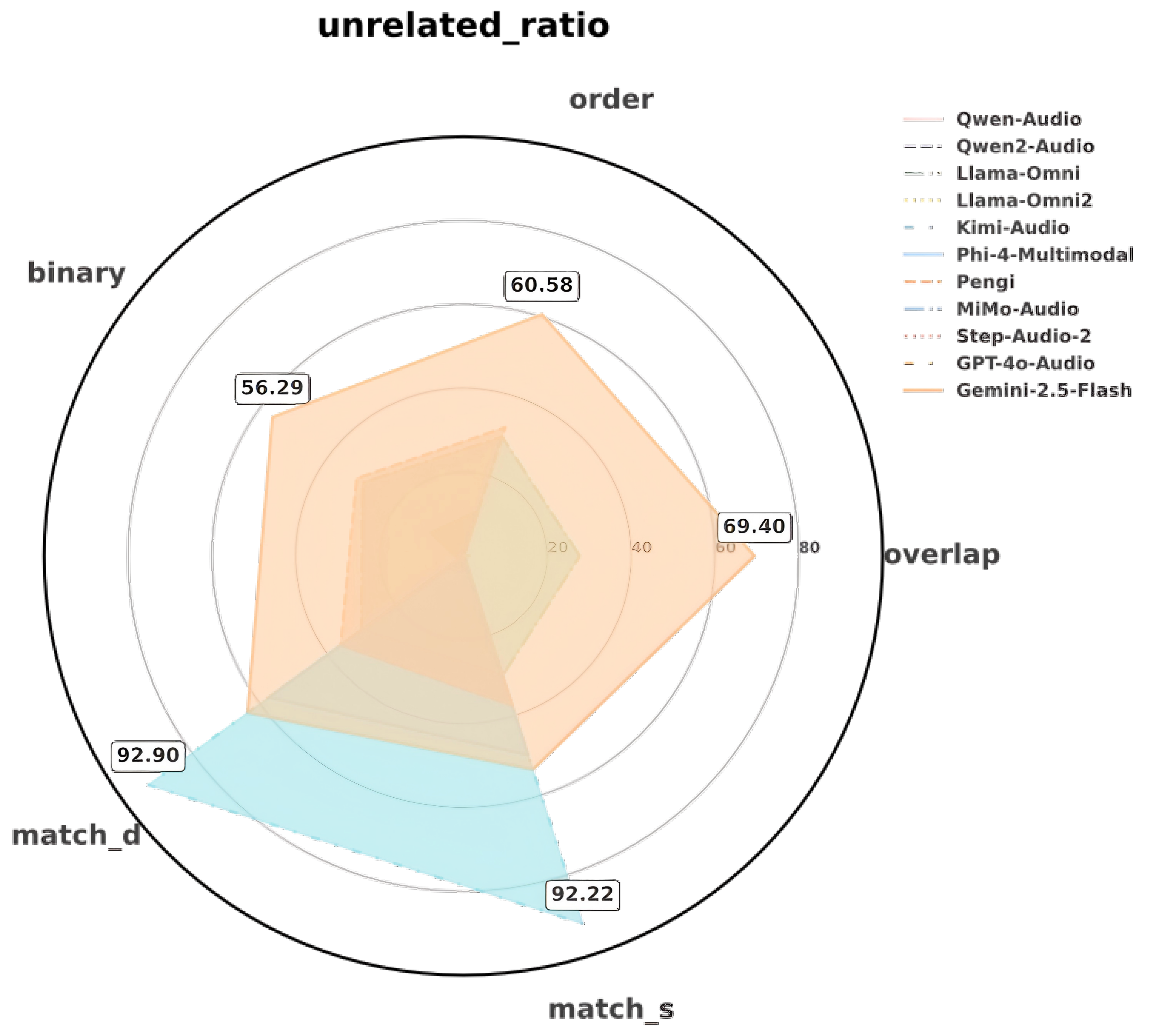} &
\includegraphics[width=0.32\textwidth]{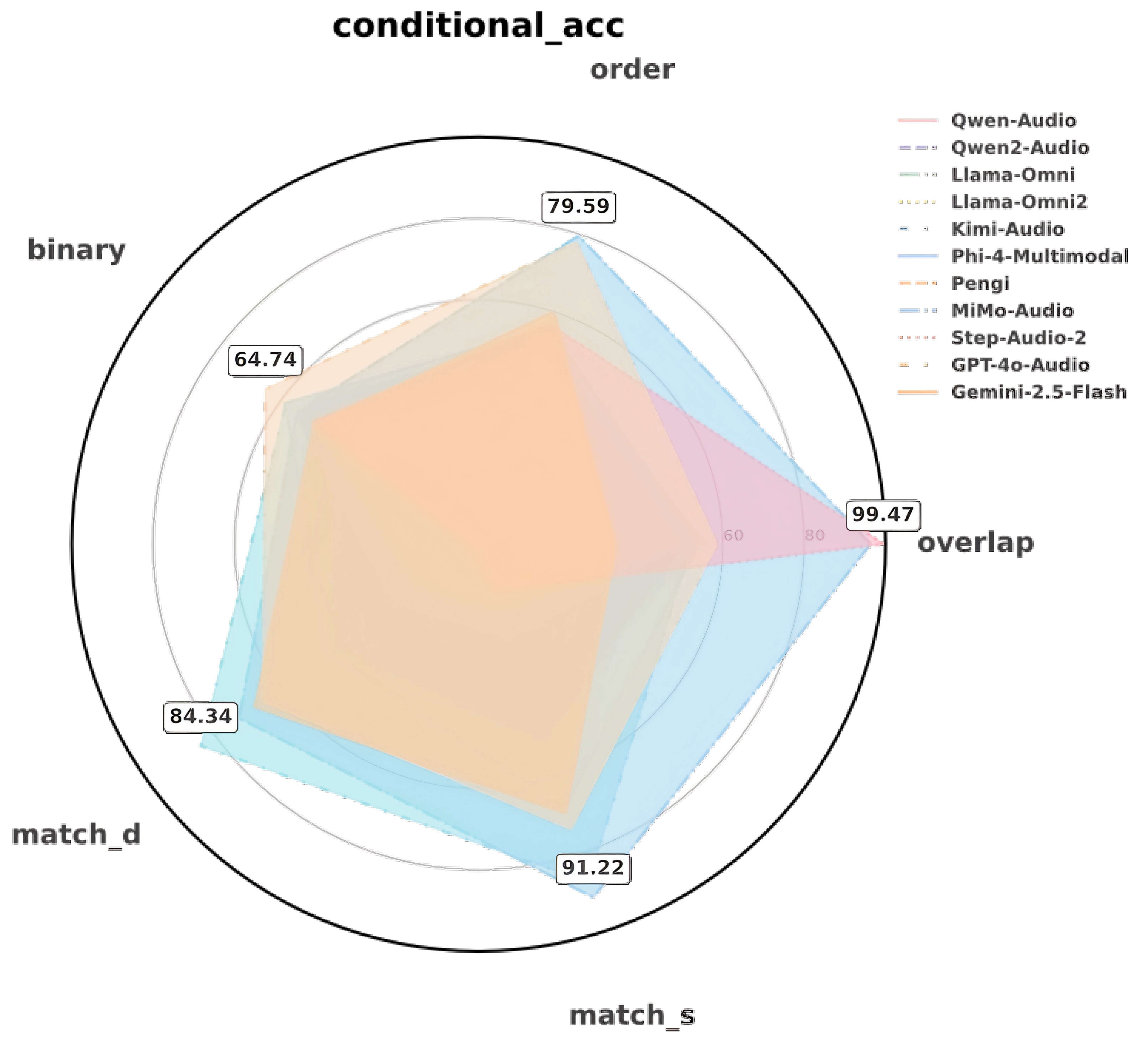} \\

\includegraphics[width=0.32\textwidth]{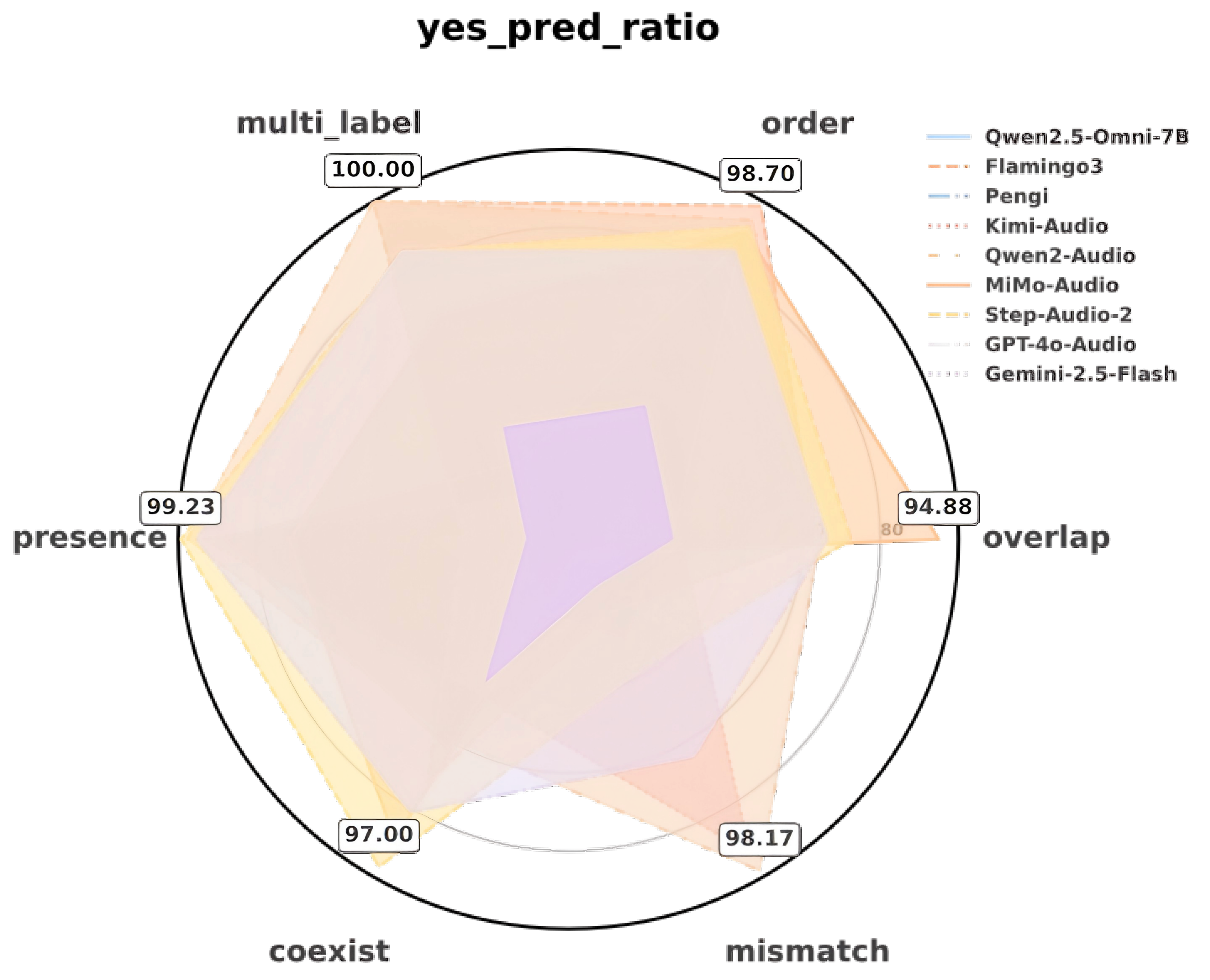} &
\includegraphics[width=0.32\textwidth]{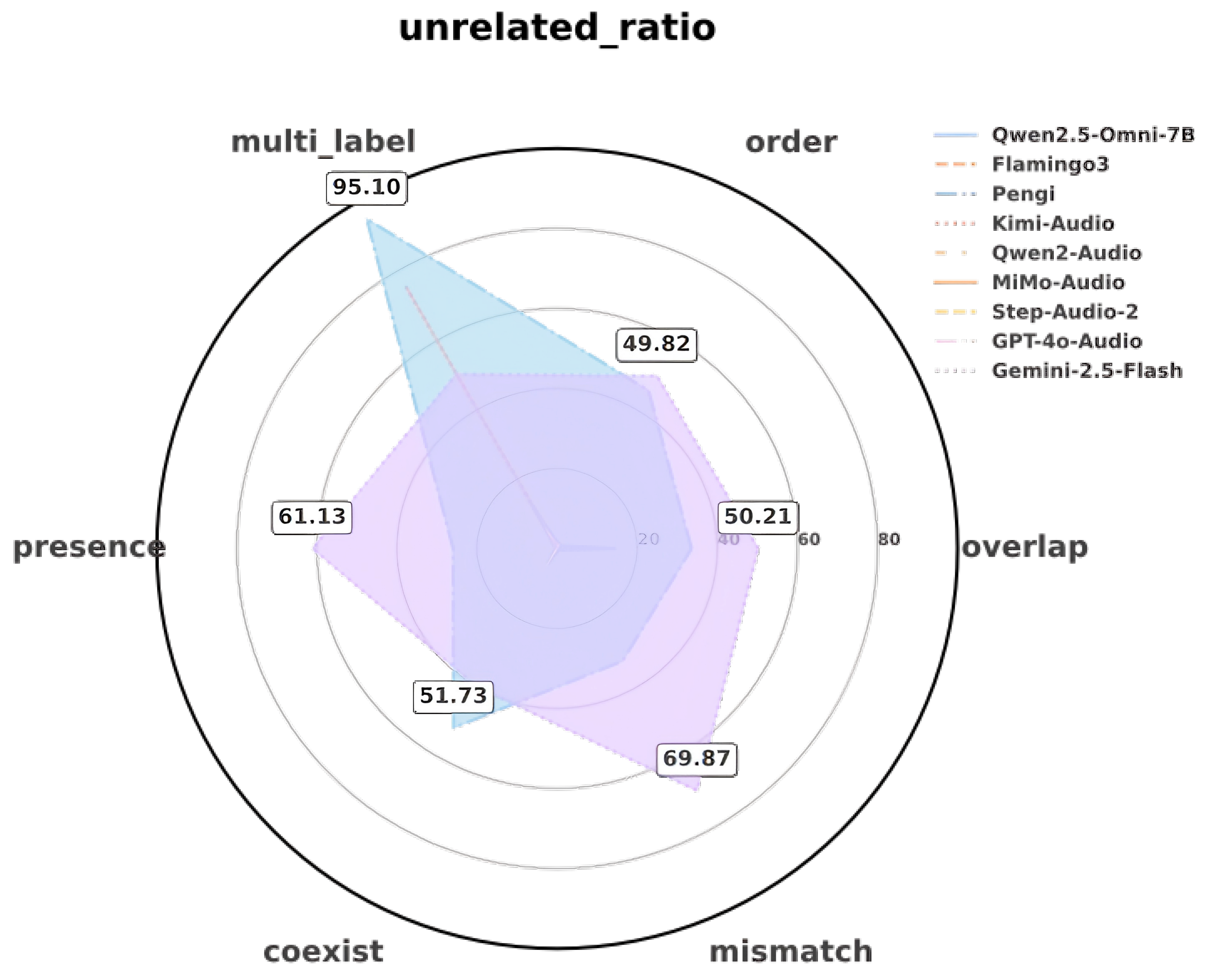} &
\includegraphics[width=0.32\textwidth]{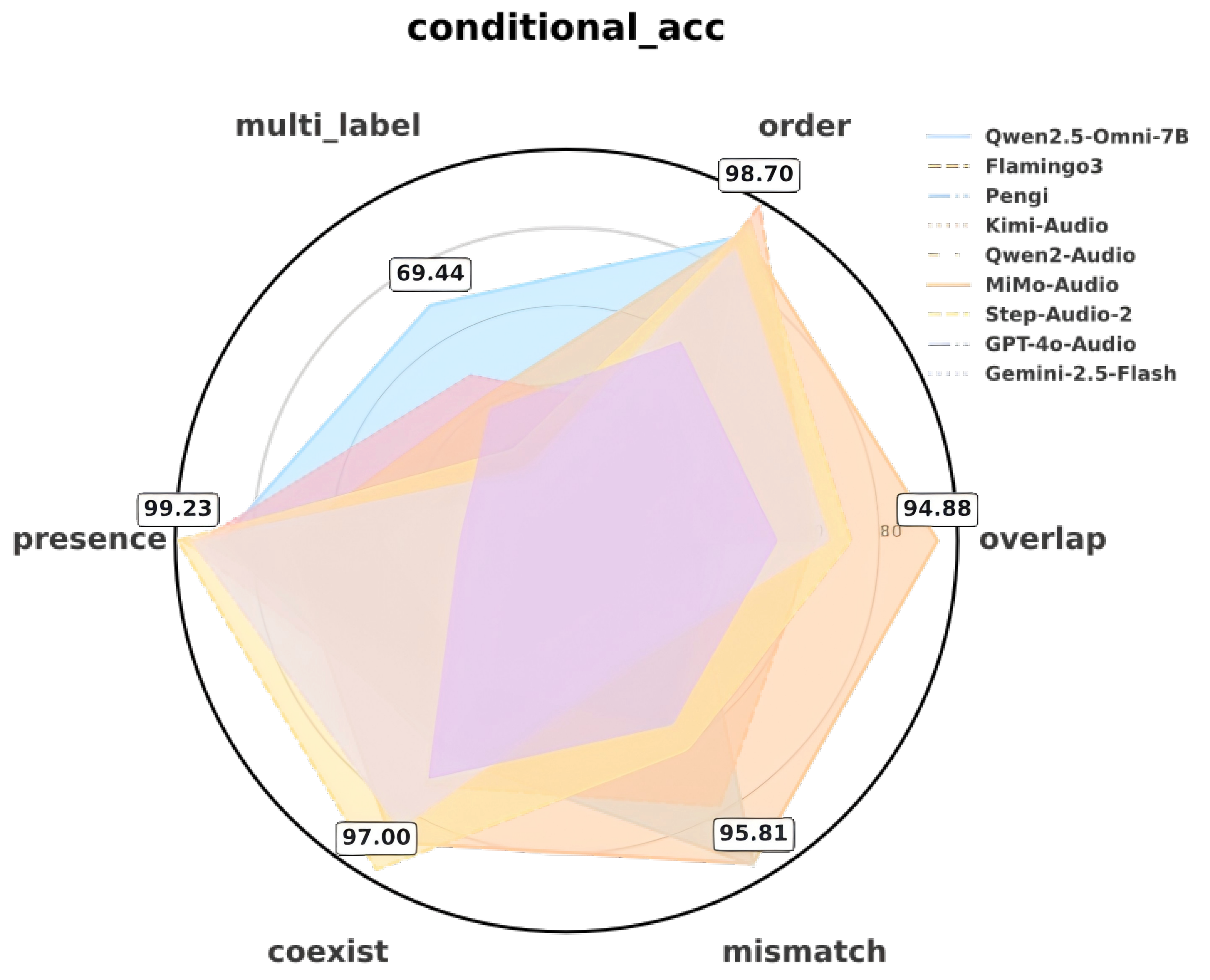} \\
\end{tabular}

\caption{
Yes/No Bias analysis of speech and environmental sound.
From left to right: Yes-pred Ratio, Unrelated Error Ratio, and Conditional Accuracy.
Higher Yes-pred Ratios combined with low conditional accuracy indicate strong affirmative bias rather than evidence-grounded binary reasoning.
}

\label{fig:yesno_bias}
\end{figure*}

\subsection{Evaluation Metrics}

We evaluate hallucination behaviors in LALMs using metrics beyond standard accuracy, capturing both response correctness and bias patterns.

\paragraph{Accuracy.} Accuracy measures the fraction of prompts with well-defined ground truth that are answered correctly by LALMs:
\begin{equation}
\small
\text{Accuracy}_d = \frac{1}{|P_d|} \sum_{p \in P_d} \mathbf{1}\{ \hat{y}_p = y_p \},
\end{equation}
where $P_d$ denotes prompts in domain $d$, $y_p$ is the ground-truth answer, and $\hat{y}_p$ is the model prediction. This allows fine-grained analysis across domains and reasoning skills. Accuracy captures explicit hallucination cases where model predictions contradict clearly defined ground-truth audio evidence.

\paragraph{Yes/No Bias Test.} 
This test diagnoses systematic bias in binary responses using three complementary measures.
% Diagnoses systematic bias in binary responses via three complementary measures:

\begin{equation}
\small
\text{Yes-p Ratio} = \frac{\sum_{p \in P_{\text{binary}}} \mathbf{1}\{\hat{y}_p = \text{Yes}\}}{|P_{\text{binary}}|},\\
\end{equation}
\begin{equation}
\small
\text{Unrelated Ratio} = \frac{\sum_{p \in P_{\text{binary}}} \mathbf{1}\{\hat{y}_p \neq y_p \wedge \hat{y}_p \}}{\sum_{p \in P_{\text{binary}}} \mathbf{1}\{\hat{y}_p \neq y_p\}},\\
\end{equation}
\begin{equation}
\small
\text{Conditional Accuracy} = \frac{\sum_{p \in P_{\text{binary}}} \mathbf{1}\{\hat{y}_p = y_p\}}{|P_{\text{binary}}|},
\end{equation}
where $P_{\text{binary}}$ is the set of Yes/No prompts, and $p \in P_{\text{binary}}$ splits by predicted class. These measures diagnose systematic affirmative bias, where models tend to produce unsupported positive responses despite insufficient or contradictory evidence, a common form of hallucination behavior.

\paragraph{False Refusal Rate (FRR).} FRR captures cases where LALMs abstain despite a valid prompt:
\begin{equation}
\small
\text{FRR}_d = \frac{|\{ p \in P_d : \hat{y}_p \in \mathcal{R} \}|}{|P_d|},
\end{equation}
where $\mathcal{R}$ denotes refusal responses. FRR reflects over-conservative hallucination behavior, where models fail to respond despite the presence of sufficient evidence, indicating breakdowns in evidence-based decision making. 

\subsection{Results}

\paragraph{Accuracy Analysis.}
Tables \ref{speech_accuracy}, \ref{audio_accuracy}, and \ref{music_accuracy} report classification accuracy on \textit{HalluAudio} across the speech, environmental sound, and music domains. Bold and underlined values indicate the maximum and minimum.
Overall, hallucination behavior is highly task- and domain-dependent, and no single model exhibits uniformly robust performance.
% Overall, hallucination behaviors remain highly task- and domain-dependent, with no single model demonstrating uniformly robust performance.

In the \textit{speech domain}, structural hallucinations strongly impact temporal tasks such as counting and ordering. MiMo-Audio and Step-Audio-2 perform well, while Qwen-Audio, Llama-Omni, and Pengi remain below 50\% on several subtasks. Semantic hallucinations appear in transcription-related prompts: Phi-4-Multimodal degrades under noise and gender perturbations, and Kimi-Audio shows inconsistent recognition. GPT-4o-Audio is more balanced overall but still struggles with fine-grained perceptual judgments.

In the \textit{sound domain}, perceptual and structural hallucinations are amplified in multi-label, coexistence, and adversarial-negative settings. MiMo-Audio and Qwen2.5-Omni consistently outperform others on temporal comparison tasks, while Audio Flamingo3 and Pengi struggle with multi-label sound recognition. Loudness comparison remains unstable across models, and random-false prompts expose divergent behaviors: some models confidently hallucinate sound presence, while others over-refuse despite clear acoustic evidence.

In the \textit{music domain}, models frequently exhibit semantic hallucinations in genre and instrument identification under ambiguous audio. GPT-4o-Audio and MiMo-Audio perform relatively well, while Pengi and Qwen2-Audio approach near-random accuracy. Structural and temporal errors persist in stroke and tonic counting, with MiMo-Audio leading and Qwen2-Audio and Gemini-2.5-Flash falling below 15\%. Perceptual hallucinations also occur in single- vs. multi-source detection, where Qwen2.5-Omni and GPT-4o-Audio near perfect accuracy, unlike several open-source models.

Across the three domains, both open-source and proprietary LALMs show distinct, non-uniform hallucination patterns, highlighting the need for fine-grained cross-domain evaluation beyond aggregate accuracy.

\begin{figure}[htbp]
    \centering
    \includegraphics[width=0.49\textwidth]{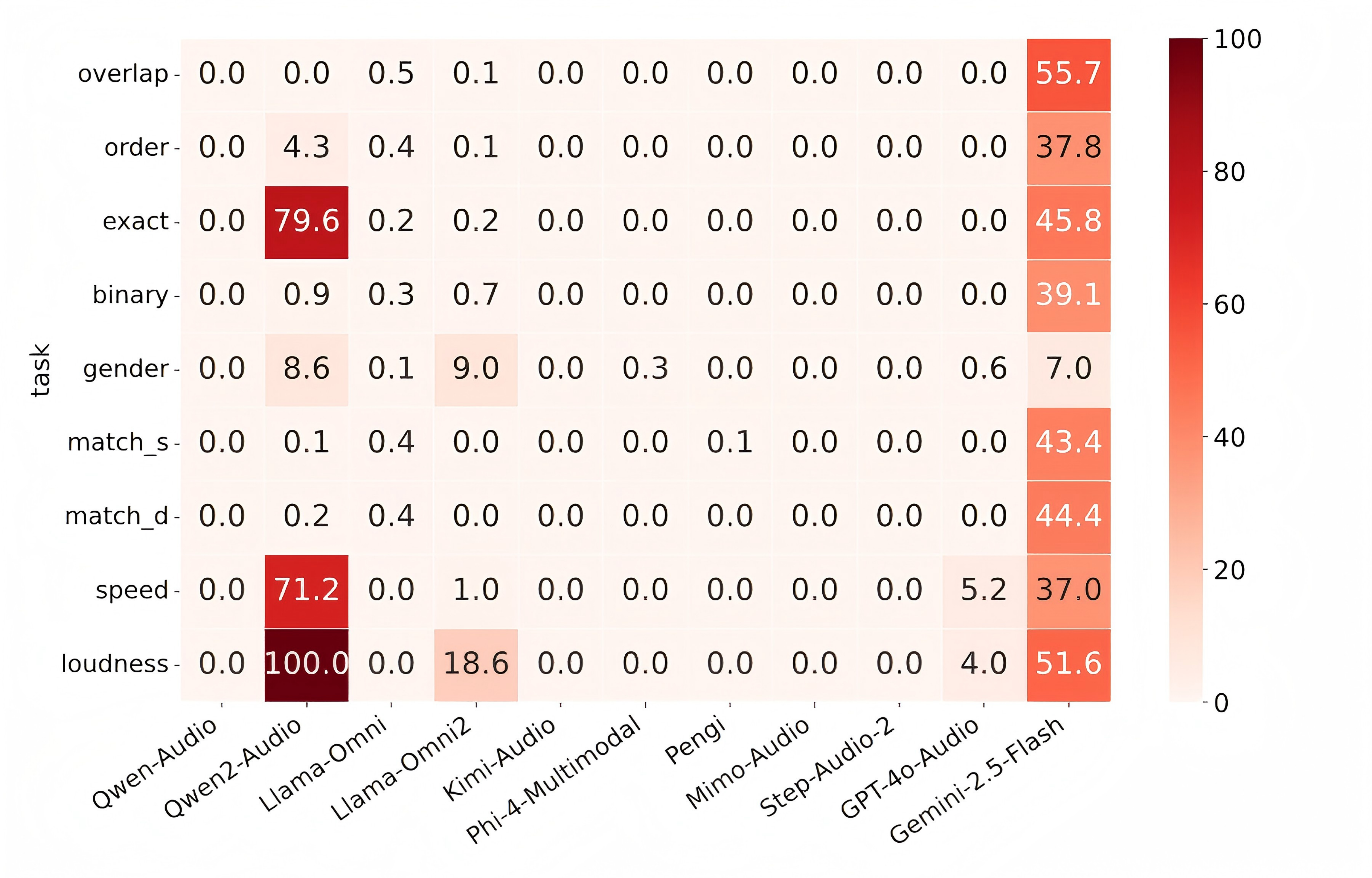}
    \hfill
    \includegraphics[width=0.49\textwidth]{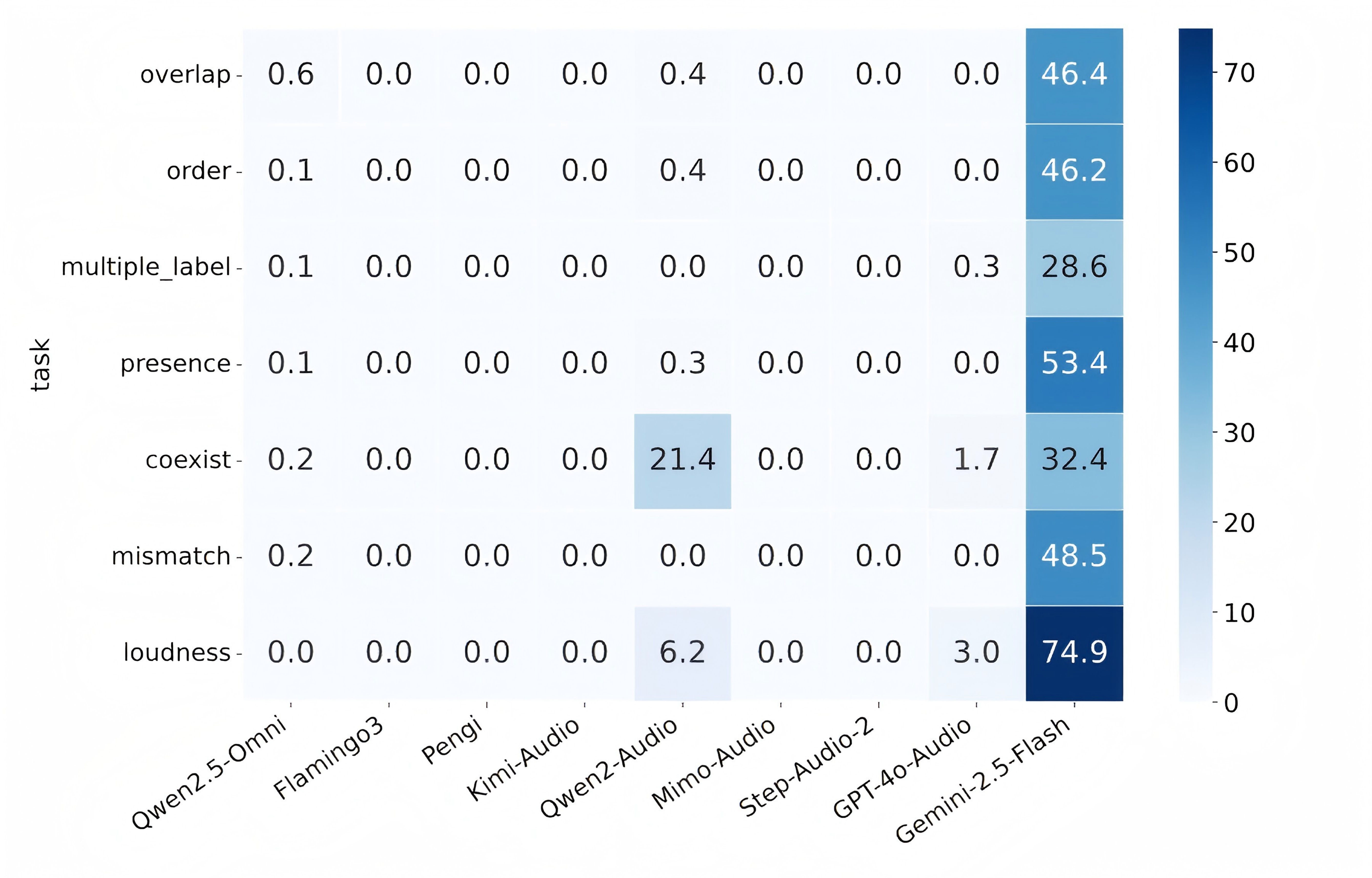}
    \caption{False Refusal Rate in speech and environmental sound domains. Each heatmap reports task-specific refusal frequencies for different models, where higher values indicate a stronger tendency to refuse answering despite the presence of a valid ground-truth answer.}
    \label{fig:false_refusal}
\end{figure}

\paragraph{Yes/No Bias Analysis.} Figure~\ref{fig:yesno_bias} examines Yes/No bias from three complementary views. The \emph{Yes prediction ratio} shows that Qwen2.5-Omni, Qwen2-Audio, and Kimi-Audio consistently over-predict affirmative answers across speech and environmental sound, with the strongest skew on ordering and counting tasks, indicating a largely domain-agnostic bias. 

The \emph{unrelated error ratio} suggests that affirmative bias does not necessarily translate into semantically unrelated errors: Qwen-series models maintain relatively low unrelated ratios, while Pengi and Audio Flamingo3 deteriorate sharply under swap and deletion perturbations, reflecting weakened grounding. 

Finally, \emph{conditional accuracy} reveals asymmetric decision behavior—models with strong affirmative bias achieve higher accuracy on positive cases but underperform on negatives, whereas Phi-4-Multimodal exhibits a more balanced yet conservative profile. Overall, affirmative bias is most pronounced in environmental sound tasks, where negative evidence must be inferred from the absence of acoustic events.
Detailed analysis of music domain can be found in the appendix \ref{music}.

\paragraph{False Refusal Analysis.} Figure~\ref{fig:false_refusal} reports false refusal behaviors in speech and environmental audio domains.
The behavior of music domain can be found in the appendix \ref{music}.
In the \emph{speech domain}, false refusals concentrate on structurally demanding tasks such as counting, speed, and loudness comparison. Qwen2-Audio shows extreme refusal on count and speed, while Gemini-2.5-Flash exhibits consistently high refusal across most tasks, indicating over-conservative abstention. In contrast, Phi-4-Multimodal, Kimi-Audio, MiMo-Audio, and Step-Audio-2 maintain near-zero refusal, suggesting stronger grounding under clear acoustic evidence.

In \emph{environmental sound domain}, refusal behavior reflects perceptual uncertainty rather than task complexity. Qwen2-Audio spikes on coexist, and loudness, and Gemini-2.5-Flash again refuses broadly, whereas Qwen2.5-Omni, MiMo-Audio, and Step-Audio-2 remain stable even under adversarial-negative settings. 

Overall, false refusals are not uniform safety behaviors but a distinct hallucination mode, arising from failures in structural reasoning, perceptual confidence, or overly conservative decision policies, with Gemini-2.5-Flash and Qwen2-Audio exhibiting the most severe over-refusal tendencies.

Overall, false refusals form a distinct hallucination mode, driven by reasoning, perceptual, or conservative policy failures, with Gemini-2.5-Flash and Qwen2-Audio showing the most severe cases.

\subsection{Qualitative Comparison}

Across domains, models exhibit distinct and often inconsistent hallucination profiles, indicating strong domain- and structure-dependence in LALMs. In the \emph{speech domain}, performance is stable on basic recognition but degrades sharply on structurally demanding tasks such as counting, ordering, and speed comparison. Qwen2-Audio and Gemini-2.5-Flash frequently default to refusal, whereas Phi-4-Multimodal and MiMo-Audio maintain more robust structural grounding. In contrast, Kimi-Audio tends to over-assert on binary and invalid queries, producing confident yet weakly supported responses.

In the \emph{environmental sound domain}, perceptual hallucinations dominate. Qwen2.5-Omni and MiMo-Audio show more balanced behavior on event presence and co-occurrence, while Audio Flamingo3 and Pengi degrade under multi-label and adversarial-negative settings. Several models exhibit affirmative bias without sufficient grounding, highlighting weakness in absence-based reasoning.

The \emph{music domain} shows the greatest model divergence. GPT-4o-Audio and MiMo-Audio are more robust on high-level semantic queries, while most models struggle with structural and temporal reasoning such as counting and order. Qwen2-Audio and Gemini-2.5-Flash exhibit near-random performance or abrupt refusal spikes, revealing fragile musical structure understanding. Overall, no model is consistently robust across domains and hallucination types, highlighting the need for multi-domain, multi-metric evaluation of LALMs.

\section{Conclusion}

This paper presents HalluAudio, a diagnostic benchmark for systematically evaluating hallucination behaviors in LALMs across speech, environmental sound, and music domains. Through fine-grained accuracy analysis, Yes/No bias testing, and FRR evaluation, we demonstrate that hallucinations in audio understanding manifest in diverse and previously underexplored forms beyond incorrect predictions. Our findings highlight the limitations of accuracy-centric evaluation and underscore the need for reliability-oriented benchmarks to guide the development of more robust LALMs.

\newpage
% \section*{Limitations}
% This work provides a comprehensive LALMs hallucination benchmark encompassing binary classification and open-ended question-answering tasks across three domains: speech, environmental sound, and music. However, our focus remains primarily on the binary classification task, with a slightly smaller number of open-ended question-answer pairs compared to the binary classification task. Furthermore, this benchmark only includes English data; while we do not cover multilingual illusions, our methods and benchmark design are equally applicable to other languages within LALMs.

% \section*{Acknowledgments}

% This work was supported in part by the National Key Research and Development Program of China under Grant 2023YFB2603902, in part by the Major Science and Technology Specific Project of Xining under Grant 2024-Z-7.

\bibliography{main}

\newpage

\appendix

\section{Dataset Details}

\subsection{Annotator Details}

HalluAudio is constructed through a rigorous human-in-the-loop annotation and validation pipeline designed to ensure label reliability at scale. Candidate QA pairs are first generated automatically via programmatic rules and filtering procedures, after which stratified subsets are manually inspected by annotators with experience in speech and audio understanding. Each instance is reviewed across multiple independent passes to verify audio–question consistency, ground-truth correctness, and the validity of hallucination-triggering conditions. 

To quantify annotation reliability, we compute inter-annotator agreement using Cohen’s \(\kappa\). The overall agreement is 0.91, with domain-level agreement above 0.89 across speech, sound, and music subsets. Approximately 4–5\% of candidate QA pairs were revised or discarded due to annotator disagreement. Final labels are determined through majority agreement with additional review by a senior annotator when necessary.

\subsection{Dataset Construction Pipeline}

\begin{table}[htbp]
\centering
\small
\setlength{\tabcolsep}{6pt}
\renewcommand{\arraystretch}{1.2}
\begin{tabularx}{\columnwidth}{m{2.2cm} X}
\hline
Field & Description \\
\hline
\texttt{Question} &
The natural language question generated from a task-specific prompt template, describing the perceptual or semantic judgment to be made based on the input audio. \\

\texttt{Answer} &
The ground-truth answer associated with the question. For binary tasks, the answer is \texttt{Yes} or \texttt{No}, while other tasks use predefined task-specific answer spaces. \\

\texttt{Fname} &
The unique identifier or filename of the audio clip used to construct the task instance, enabling traceability to the original audio source. \\

\texttt{Audio} &
The raw audio content corresponding to the task instance, stored as serialized audio bytes and provided directly to the model during inference. \\
\hline
\end{tabularx}
\caption{Data instance structure in the HalluAudio dataset.}
\label{tab:data_structure}
\end{table}

The HalluAudio dataset is constructed through a unified and systematic pipeline that transforms curated audio recordings into structured audio--question pairs for hallucination evaluation. For each selected audio clip, we generate multiple task instances by pairing the audio with parameterized prompt templates corresponding to different perceptual and reasoning tasks.

Each dataset instance follows a consistent structure, consisting of a natural language question, its ground-truth answer, a reference to the source audio file, and the raw audio content itself. This design ensures that all model inputs are grounded directly in the audio signal rather than intermediate representations or extracted features. Table~\ref{tab:data_structure} summarizes the fields contained in each data instance.

Task instances are created by instantiating template variables using verified annotations associated with each audio clip, such as transcriptions, sound event labels, or musical attributes. Valid queries are guaranteed to be answerable from the audio evidence alone, while invalid or unanswerable queries are intentionally constructed by referencing attributes absent from the audio. All instances are automatically validated to ensure consistency between the question, answer space, and audio content.

\subsection{Prompt Template Design}

Prompt templates in HalluAudio are designed to generate the \texttt{Question} field of each dataset instance, probing hallucination behaviors under diverse linguistic and perceptual conditions while preserving consistent task semantics. For each task category, multiple templates are constructed with varied phrasing styles and contextual formulations to reduce sensitivity to prompt-specific artifacts.

In the speech domain, prompt templates target fine-grained acoustic and semantic reasoning, including word occurrence, temporal ordering, counting, and speaker-related attributes. In the environmental sound domain, templates focus on sound event perception and co-occurrence reasoning, such as sound presence, overlap detection, and multi-label identification. In the music domain, prompts emphasize perceptual and categorical judgments, including instrument identification, genre matching, loudness comparison, and speech--music discrimination.

\begin{figure*}[t]
\centering
\setlength{\tabcolsep}{2pt}

\begin{tabular}{ccc}

\includegraphics[width=0.32\textwidth]{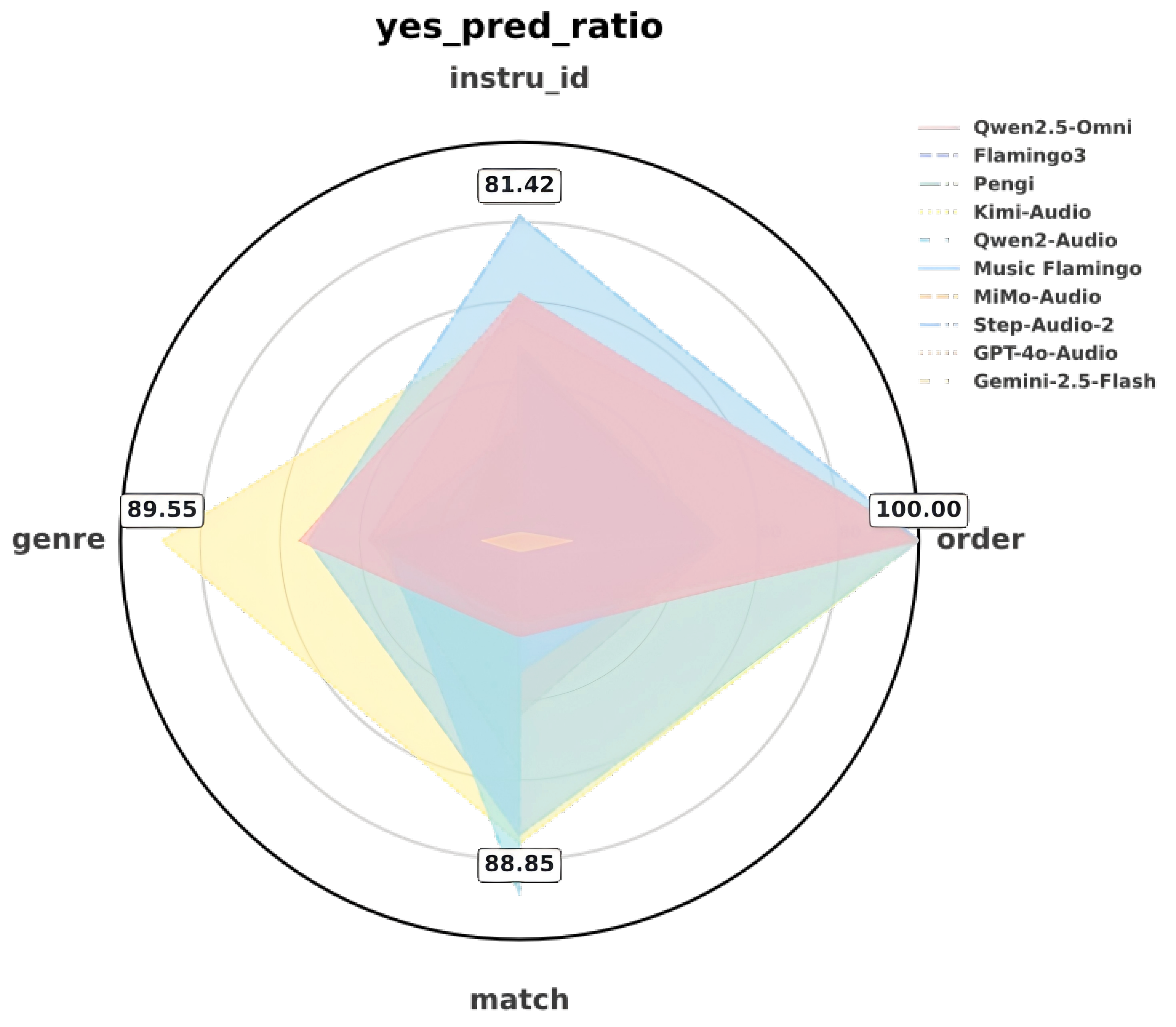} &
\includegraphics[width=0.32\textwidth]{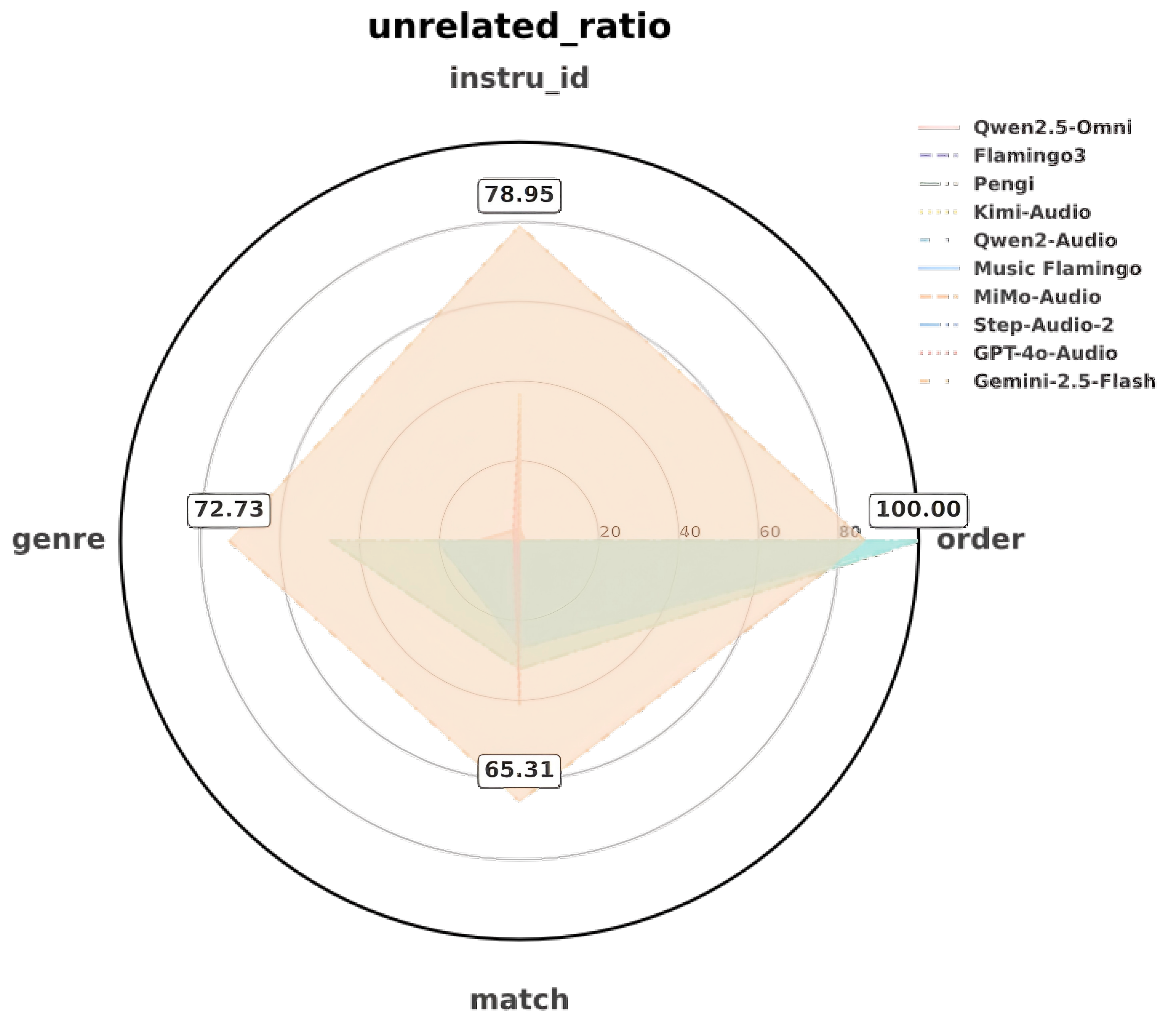} &
\includegraphics[width=0.32\textwidth]{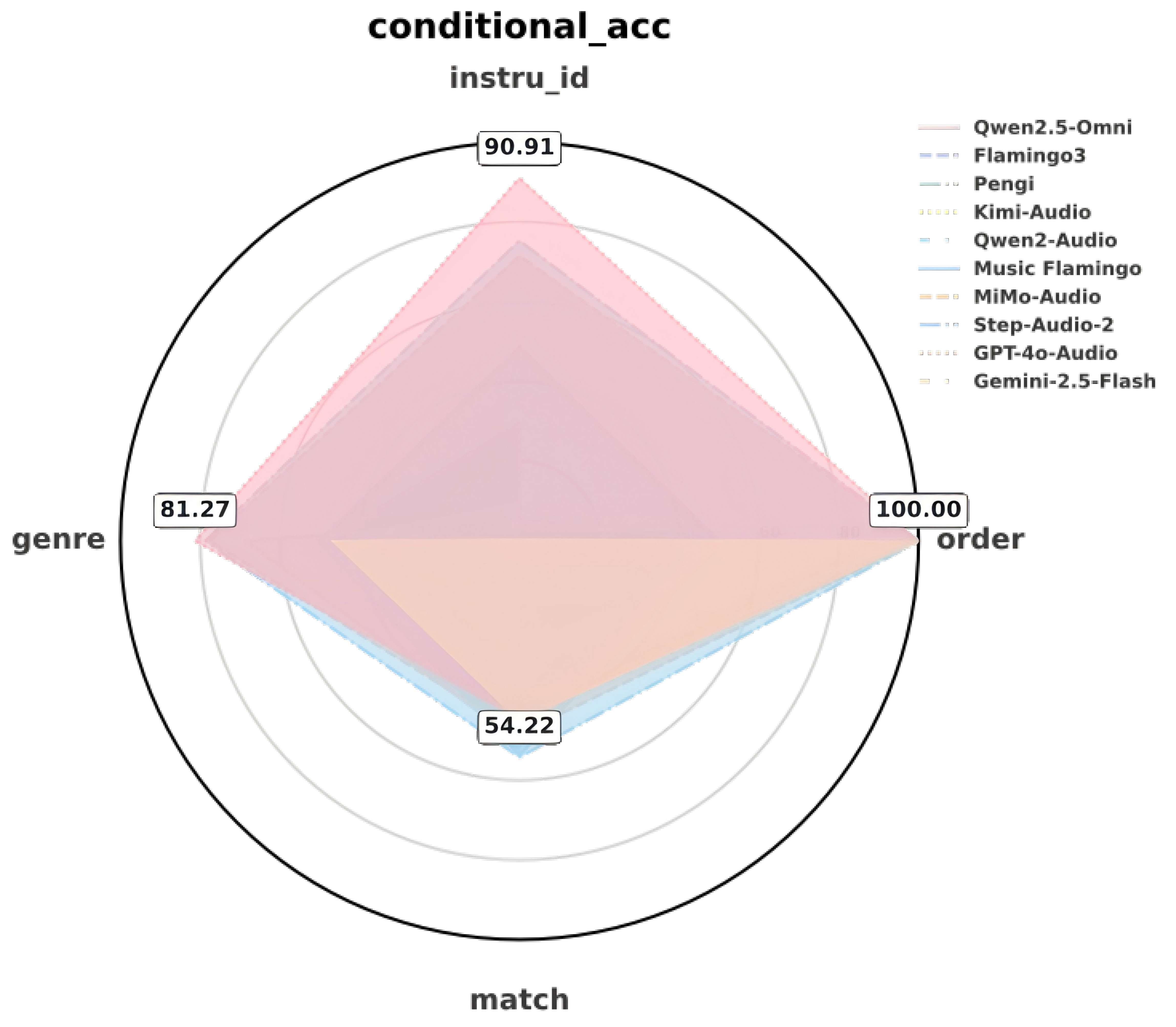}
\end{tabular}

\caption{
Yes/No Bias Analysis in music domain.
From left to right: Yes-pred Ratio, Unrelated Error Ratio, and Conditional Accuracy.
Higher Yes-pred Ratios combined with low conditional accuracy indicate strong affirmative bias rather than evidence-grounded binary reasoning.
}

\label{fig:music_bias}
\end{figure*}

Across all domains, both valid and intentionally invalid prompt templates are included. Valid prompts correspond to answerable questions grounded in the audio content, while invalid prompts are constructed to be unanswerable by design, serving as controlled probes for hallucinated responses and refusal behaviors. All templates are parameterized and instantiated automatically, ensuring scalable and reproducible generation of question--audio pairs.

\section{Music data analysis}

\label{music}

\subsection{Yes/No Bias Analysis}

In Figure \ref{fig:music_bias}, the Yes/No bias analysis reveals pronounced asymmetry across task types and models.
The \emph{Yes prediction ratio} shows that several models, notably Qwen2.5-Omni and GPT-4o-Audio, exhibit near-saturated affirmative responses on ordering-related queries, indicating a strong tendency to default to positive judgments when reasoning about musical structure. Genre and instrument identification tasks display more dispersed behavior, suggesting comparatively better calibration under high-level semantic cues, while matching tasks expose substantial variability across models.

The \emph{unrelated error ratio} highlights that high affirmative bias does not necessarily translate to semantically grounded errors. Ordering tasks consistently induce the highest unrelated responses, implying that structural musical reasoning is particularly vulnerable to hallucinated content rather than simple misclassification. In contrast, genre-related queries show lower unrelated ratios, reflecting stronger alignment between model responses and the intended decision space.

Finally, \emph{conditional accuracy} exposes a clear imbalance between positive and negative cases. Models with strong affirmative tendencies achieve high accuracy on positive instances but degrade sharply on negative matching tasks, revealing limited sensitivity to absence or contradiction in musical evidence. Overall, these results indicate that hallucination in music-centric tasks is tightly coupled with structural reasoning demands, and that high-level semantic understanding alone is insufficient to ensure reliable binary judgment in complex musical contexts.

\subsection{False Refusal Analysis.}

\begin{figure}[htbp]
    \centering

    \includegraphics[width=0.45\textwidth]{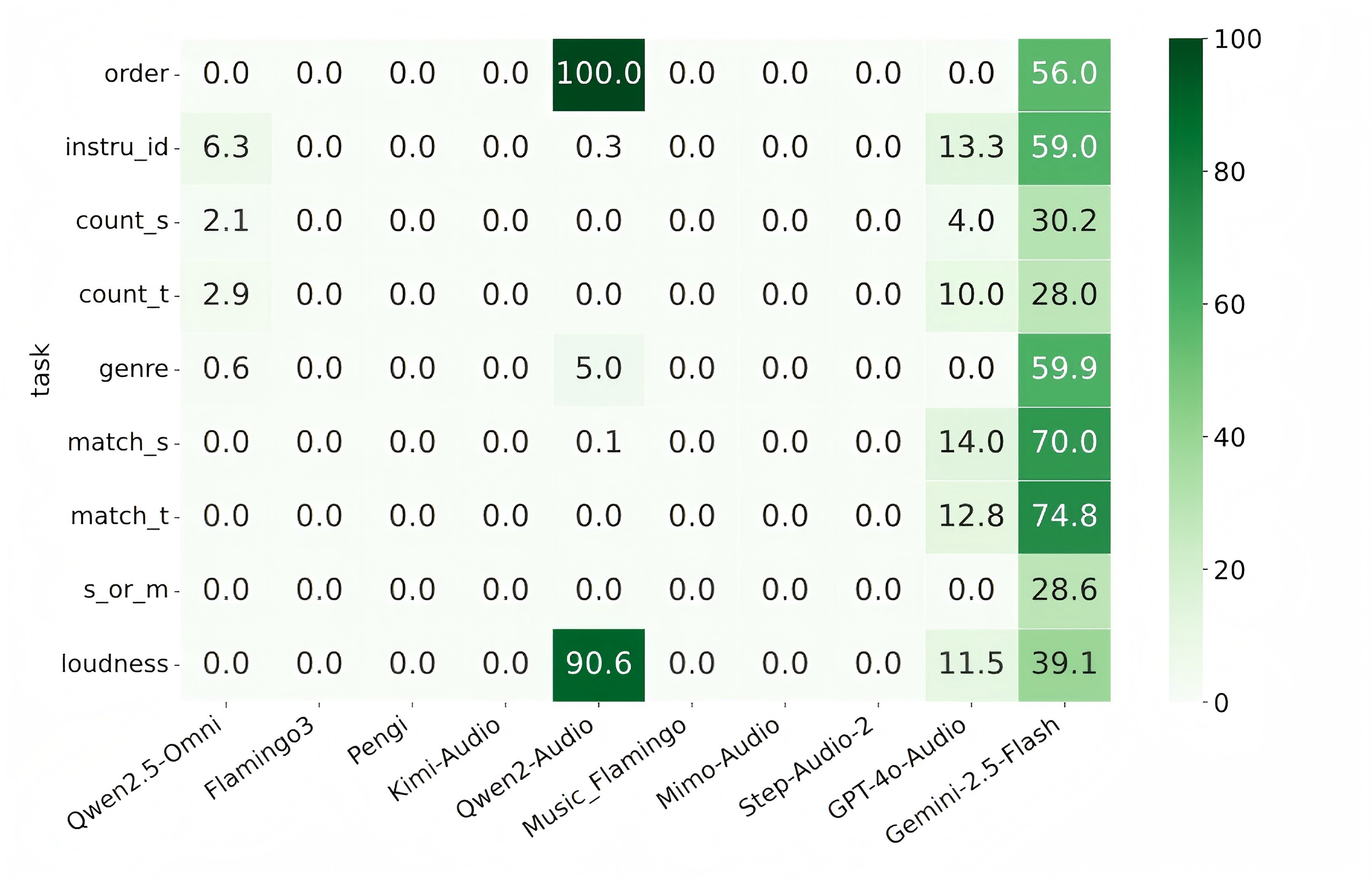}
    \caption{False Refusal Rate in music domain. Higher values indicate a stronger tendency to refuse answering despite the presence of a valid ground-truth answer.}
    \label{fig:music_refusal}
\end{figure}

In Figure \ref{fig:music_refusal}, false refusals are sparse overall but highly task-specific: Qwen2-Audio collapses on order and loudness, while Gemini-2.5-Flash shows elevated refusal across genre, counting, and matching tasks.

\section{Post-hoc Robustness Test}

\begin{table}[htbp]
\centering
\small
\begin{tabular}{lccc}
\toprule
\textbf{Model} & \textbf{Orig.} & \textbf{Para.} & \textbf{Diff.} \\
\midrule
Qwen2-Audio-7B & 50.1 & 49.9 & 0.2 \\
Kimi-Audio-7B & 48.4 & 47.9 & 0.5 \\
Pengi & 22.7 & 20.9 & 1.8 \\
MiMo-Audio-7B-Instruct & 61.3 & 60.8 & 0.5 \\
Step-Audio-2-mini & 56.8 & 56.5 & 0.3 \\
\midrule
\textbf{Average} & 47.9 & 47.2 & 0.7 \\
\bottomrule
\end{tabular}
\caption{Robustness(\%) of different audio-language models under semantic paraphrasing. Results are averaged over 1,000 sampled QA pairs across all domains.}
\label{tab:paraphrase_models}
\end{table}

To evaluate potential linguistic bias introduced by template-based prompt generation, we conduct a post-hoc robustness analysis using semantically equivalent paraphrases across multiple audio-language models.

We randomly sample 1,000 QA pairs from HalluAudio, covering speech, environmental sound, and music domains. For each instance, we manually construct a paraphrased variant that preserves semantic meaning while altering surface wording and syntactic structure.

We evaluate both original and paraphrased prompts on five representative models, including Qwen2-Audio-7B, Kimi-Audio-7B, Pengi, MiMo-Audio-7B-Instruct, and Step-Audio-2-mini, under identical inference settings. As shown in Table~\ref{tab:paraphrase_models}, all models exhibit minimal performance variation under paraphrasing. The absolute difference remains consistently small, ranging from 0.2\% to 1.8\%, with an average deviation of 0.7\%.

This consistency across diverse architectures indicates that model behavior is largely invariant to superficial linguistic variations. The results suggest that performance differences are driven primarily by task structure and acoustic reasoning requirements rather than prompt wording.

Overall, these findings provide strong evidence that the template-based generation strategy in HalluAudio does not introduce significant linguistic bias, and that the benchmark reliably reflects hallucination behavior instead of prompt sensitivity artifacts.

\section{LALMs Participating in Benchmark}

Based on the different focuses of different LALMs, we selected different LALMs for different domains, and the detailed data is listed in Table \ref{tab: LALMs}.

\begin{table}[htbp]
\centering
\small
\setlength{\tabcolsep}{3.5pt}
\begin{tabular}{l c c c c}
\toprule
\textbf{Model} & \textbf{Speech} & \textbf{Sound} & \textbf{Music} & \textbf{Source} \\
\midrule
Qwen-Audio-Chat      & \checkmark &              &              & Open \\
Qwen2-Audio-7B       & \checkmark & \checkmark   & \checkmark   & Open \\
Qwen2.5-Omni-7B      &            & \checkmark   & \checkmark   & Open \\
Llama-3.1-8B-Omni         & \checkmark &              &              & Open \\
Llama-Omni2-7B       & \checkmark &              &              & Open \\
Kimi-Audio-7B        & \checkmark & \checkmark   & \checkmark   & Open \\
Phi-4-Multimodal     & \checkmark &              &              & Open \\
Audio Flamingo-3     &            & \checkmark   & \checkmark   & Open \\
Music Flamingo       &            &              & \checkmark   & Open \\
Pengi                & \checkmark & \checkmark   & \checkmark   & Open \\
MiMo-Audio-7B-Instruct                & \checkmark & \checkmark   & \checkmark   & Open \\
Step-Audio-2-mini                & \checkmark & \checkmark   & \checkmark   & Open \\
\midrule
GPT-4o-Audio-Preview                & \checkmark & \checkmark   & \checkmark   & Closed \\
Gemini-2.5-Flash         & \checkmark & \checkmark   & \checkmark   & Closed \\
\bottomrule
\end{tabular}
\caption{LALMs evaluated in HalluAudio across different domains.}
\label{tab: LALMs}
\end{table}

\textbf{Qwen-Audio-Chat}: An open-source multimodal audio-language model by Alibaba Cloud. It extends the Qwen-Audio foundation by instruction fine-tuning to support multi-turn spoken dialogue. The model accepts diverse audio inputs along with text and generates text responses. Qwen-Audio-Chat is designed for comprehensive audio understanding – including speech reasoning, sound classification, music appreciation, and even audio-based editing – within conversational contexts.

\textbf{Qwen2-Audio-7B}: A 7-billion-parameter audio-aware LLM from Alibaba’s Qwen series. It supports two interactive modes: "voice chat" and "audio analysis". Qwen2-Audio-7B can perform tasks like ASR, audio classification, speech-to-text translation, and emotion or sound recognition across multiple languages. The system is released with an instruct-tuned variant and achieves strong benchmarks on standard speech and audio understanding tasks.

\textbf{Qwen2.5-Omni-7B}: A 7B open-source multimodal model by Alibaba. It is designed to perceive and integrate text, images, audio, and video inputs simultaneously, while generating textual and natural speech outputs in real time. Qwen2.5-Omni uses a "Thinker-Talker" architecture with a time-aligned embedding scheme to synchronize audio/video timestamps. This enables features like real-time voice and video chat. In evaluations, Qwen2.5-Omni outperforms similarly-sized single-modality models on joint tasks and exceeds the audio capabilities of Qwen2-Audio.

\textbf{LLaMA-3.1-8B-Omni}: A speech-enabled LLM built on Meta’s Llama-3.1 8B Instruct model. LLaMA-Omni integrates a pretrained speech encoder, a speech adaptor, and a streaming speech decoder with the base LLM. This design eliminates the need for intermediate transcription: it directly generates text and speech responses from spoken instructions. The result is low-latency, high-quality spoken dialogue – the model can answer and even speak back with a latency on the order of a few hundred milliseconds while maintaining content fidelity and natural style.

\textbf{LLaMA-Omni2-7B}: A 7B variant of the LLaMA-Omni2 series. Built on Qwen2.5, this model incorporates a speech encoder and an autoregressive streaming speech decoder into the LLM. It enables real-time spoken chat: given speech input, the model can generate text or speech answers on the fly. Even though it was trained on only \~200K multi-turn speech QA pairs, LLaMA-Omni2 models exhibit strong performance on spoken dialog and instruction tasks, surpassing prior speech-language models on several benchmarks.

\textbf{Kimi-Audio-7B}: A 7B open-source audio foundation model by MoonshotAI. Kimi-Audio is designed to handle a wide variety of audio tasks in one model. Its input encoder is "hybrid": incoming audio is tokenized into discrete semantic tokens and also encoded into continuous acoustic features. These features feed into a transformer LLM core, which has parallel output heads for generating text tokens and audio tokens. Kimi-Audio was pretrained on over 13 million hours of diverse audio and text, giving it strong audio-language understanding. It achieves state-of-the-art results on tasks like ASR, audio question-answering, audio captioning, emotion recognition, sound classification, and even end-to-end speech conversation.

\textbf{Phi-4-Multimodal}: A 3.8B open-source small multimodal model from Microsoft. It unifies text, vision, and speech/audio in one model using a "mixture-of-LoRAs" approach: the base language model is frozen and modality-specific LoRA adapters are added for vision and audio. Phi-4-Multimodal thus natively supports inputs like speech+text or image+audio. Despite its compact size, it achieves very strong performance on speech tasks for a model of its scale: for example, it ranks first on the open multilingual ASR leaderboard and excels at speech translation and QA. Notably, it is the first open-source model to include a speech summarization capability, highlighting its comprehensive audio understanding.

\textbf{Audio Flamingo 3}: An open-source Large Audio-Language Model by NVIDIA. AF3 advances unified audio reasoning across speech, environmental sounds, and music. It builds on Flamingo-style architecture with a unified audio encoder and adds novel features like flexible chain-of-thought reasoning and long-context comprehension. AF3 supports multi-turn audio dialogues and even voice-to-voice conversational response. In benchmarks, Audio Flamingo 3 sets new state-of-the-art scores on over 20 public audio understanding and reasoning tasks.

\textbf{Music Flamingo}: An open-source NVIDIA model specialized for music understanding. Music Flamingo analyzes complex musical audio with deep musical knowledge. It can generate rich, theory-aware captions and answers about music attributes. The model is trained with reasoning-centric methods and can process full-length songs. In evaluations it establishes new SOTA on more than 10 music-related tasks.

\textbf{Pengi}: An audio-language model by Microsoft. Pengi reframes all audio tasks as text-generation tasks. It uses an audio encoder to convert any input audio into embeddings, concatenates this with any text prompt, and feeds it into a pretrained frozen language model. This unified approach allows both open-ended tasks and closed tasks to be handled without task-specific fine-tuning. 

\textbf{MiMo-Audio-7B-Instruct}: A 7B open-source audio LLM by Xiaomi. The core MiMo-Audio model is pretrained on a massive scale, which enables emergent few-shot generalization to new audio tasks. Even without fine-tuning, the 7B base MiMo-Audio already achieves state-of-the-art open-model performance on standard speech and audio understanding benchmarks. After instruction fine-tuning, MiMo-Audio yields leading open-source results on audio comprehension, spoken dialogue, and TTS instructions. It can generalize to tasks not in its training data and produce realistic continuous speech as part of its outputs.

\textbf{Step-Audio-2-Mini}: The 8B-parameter variant of StepFun’s Step-Audio 2 family. Step-Audio 2 is an end-to-end multimodal audio-language system designed for "industry-strength" audio understanding and conversation. It integrates a latent audio encoder and uses reinforcement learning focused on reasoning. Importantly, Step-Audio-2 generates discrete audio tokens as part of its output, allowing it to capture paralinguistic cues in its responses. It also incorporates retrieval to ground its knowledge and reduce hallucinations. Trained on millions of hours of speech/audio, Step-Audio 2 achieves state-of-the-art performance on diverse audio understanding and conversational benchmarks.

\textbf{GPT-4o-Audio-Preview}: A closed-source audio-capable model from OpenAI. In preview release, GPT-4o-Audio accepts both text and audio as input and can produce either text or audio outputs. According to OpenAI’s API documentation, it supports a very large context window (128,000 tokens) for audio/text and is accessed via the Chat Completions endpoint.

\textbf{Gemini-2.5-Flash}: Google's proprietary audio LLM optimized for real-time voice interactions. The latest "Flash" version improves instruction-following and conversation smoothness for live voice agents. Gemini-2.5-Flash Native Audio can interpret complex spoken instructions, trigger external tools or function calls, and maintain natural multi-turn dialogue. Google has deployed it in products like Google Translate and in Google AI/Vertex AI for building voice agents. It supports 70+ languages in live translation, enabling seamless voice-based communication across languages.

\section{Output Normalization}

Due to the open-ended generation nature of LALMs, raw model outputs exhibit substantial variability in surface forms. For example, binary decisions may be expressed as concise tokens, extended explanations, or embedded within free-form reasoning. Such diversity poses challenges for consistent and fair evaluation across models and tasks.

To ensure reliable metric computation, we apply a text-level output normalization procedure prior to labeling. The normalization process standardizes model responses while preserving their semantic intent. Specifically, we first convert all outputs to lowercase and remove punctuation and redundant whitespace. For binary decision tasks, normalized outputs are mapped to canonical labels (\texttt{Yes} or \texttt{No}) via keyword matching. For counting tasks, numerical values are extracted from textual responses when present. In addition, responses indicating inability to answer or lack of access to audio content are explicitly mapped to a unified \texttt{Refusal} label.

Table~\ref{tab:normalization} illustrates representative examples of raw model outputs and their corresponding normalized forms.

\begin{table}[htbp]
\centering
\small
\setlength{\tabcolsep}{6pt}
\renewcommand{\arraystretch}{1.2}
\begin{tabularx}{\columnwidth}{lll}
\hline
Output Type & Raw Model Output & Normalized Form \\
\hline
Binary & Yes, it does. & Yes \\
Binary & No, it is absent. & No \\
Count & It appears three times. & 3 \\
Refusal & I cannot determine. & Refusal \\
\hline
\end{tabularx}
\caption{Examples of output normalization across different response types.}
\label{tab:normalization}
\end{table}

\begin{figure*}[t]
\centering
\begin{hallucinationcase}[Case 1: Speech -- Recognition]
\textbf{Prompt:}  
What does the speaker say in the recording?

\vspace{0.3em}
\textbf{Reference Answer:}  
None.

\vspace{0.3em}
\textbf{Erroneous Model Outputs:}
\begin{itemize}
  \item \textbf{Qwen-Audio-Chat:} The speaker says in English: ``You can see the logic in the data and the model is quite good.''
  \item \textbf{Qwen2-Audio-7B:} The speaker says ``Kids are talking by the door.''
  \item \textbf{LLaMA-3.1-8B-Omni:} Yes, I can help you with that. What would you like to know about the weather?
  \item \textbf{LLaMA-Omni2-7B:} I'm sorry, I don't have access to the recording.
  \item \textbf{Kimi-Audio-7B:} I don't know.
  \item \textbf{Phi-4-Multimodal:} The speaker says ``and subscribe for more videos!''.
  \item \textbf{Pengi:} static.
  \item \textbf{MiMo-Audio-7B-Instruct:} The speaker says ``Thank you for watching!''.
  \item \textbf{Step-Audio-2-Mini:} There is white noise. \textit{(Correct)}
  \item \textbf{GPT-4o-Audio-Preview:} ``Currently, higher education is facing a crisis with rising tuition costs\ldots''
  \item \textbf{Gemini-2.5-Flash:} ``She asked if I had seen the movie.''
\end{itemize}

\vspace{0.3em}
\textit{This case demonstrates a semantic hallucination, where models fabricate linguistic content despite the absence of intelligible speech in the audio.}
\end{hallucinationcase}
\caption{Semantic hallucination examples in speech recognition tasks.}
\label{Affirmative}
\end{figure*}

\section{Representative Failure Cases}

For each category, we report qualitative examples where multiple LALMs are evaluated on the same audio--question pair, highlighting systematic hallucination behaviors that are not fully captured by aggregate metrics.

\subsection{Hallucinated Affirmative Responses}

This category captures cases where models produce confident affirmative or content-bearing responses despite the absence of supporting acoustic evidence. As illustrated in Figure \ref{Affirmative}, when presented with an audio segment containing no intelligible speech, many LALMs nonetheless generate fluent and specific transcriptions, ranging from plausible everyday utterances to long-form, domain-irrelevant statements. These responses are not random noise but linguistically well-formed, indicating that models default to language priors rather than grounding their outputs in the actual audio input.

Such hallucinated affirmative behaviors are especially prevalent in open-ended recognition and description tasks, where the absence of explicit negative supervision encourages over-generation. While some models appropriately abstain or acknowledge uncertainty, others fabricate detailed content with high confidence, revealing a systematic affirmative bias under perceptual uncertainty. This failure mode highlights a critical reliability gap in current LALMs: when acoustic evidence is weak or missing, models often prefer generating a plausible answer over producing a grounded refusal or null response.

\subsection{Incorrect Task-grounded Answers}

\begin{figure*}[t]
\centering
\begin{hallucinationcase}[Case 2: Speech -- Temporal Comparison]
\textbf{Prompt:}  
Do the two speakers’ voices overlap in the recording?

\vspace{0.3em}
\textbf{Reference Answer:}  
Yes.

\vspace{0.3em}
\textbf{Erroneous Model Outputs:}
\begin{itemize}
  \item \textbf{Qwen-Audio-Chat:} No, the two speakers' voices do not overlap in the recording.
  \item \textbf{Qwen2-Audio-7B:} Yes, the two speakers' voices overlap in the recording as they both speak from time 0.00 to 2.39 and then from time 2.47 to 4.95. \textit{(Correct)}
  \item \textbf{LLaMA-3.1-8B-Omni:} No, she did not live with her grandfather.
  \item \textbf{LLaMA-Omni2-7B:} Yes, the two speakers' voices do overlap in the recording. \textit{(Correct)}
  \item \textbf{Kimi-Audio-7B:} no.
  \item \textbf{Phi-4-Multimodal:} The audio clip does not provide enough information to determine if the two speakers' voices overlap.
  \item \textbf{Pengi:} yes. \textit{(Correct)}
  \item \textbf{MiMo-Audio-7B-Instruct:} Yes. \textit{(Correct)}
  \item \textbf{Step-Audio-2-Mini:} No, the two speakers' voices do not overlap in the recording.
  \item \textbf{GPT-4o-Audio-Preview:} No, the two speakers' voices do not overlap in the recording.
  \item \textbf{Gemini-2.5-Flash:} As an AI, I don't have access to external audio files to listen to them. Therefore, I cannot tell you if the two speakers' voices overlap in the recording.
\end{itemize}

\vspace{0.3em}
\textit{This case exemplifies an \textbf{incorrect task-grounded answer}, where models attempt to perform the instructed task but produce answers that contradict the actual audio content. Unlike hallucinated affirmatives, the responses are relevant to the task but factually inaccurate. Such errors are particularly common in binary or multiple-choice classification tasks, where the model may overgeneralize, misinterpret subtle temporal cues, or rely on heuristics that do not align with the audio evidence.}
\end{hallucinationcase}

\caption{Incorrect task-grounded responses in a temporal comparison task, illustrating models’ tendencies to produce task-relevant but factually inconsistent outputs.}
\label{TemporalComparison}
\end{figure*}

In this category, models attempt to perform the instructed task, such as detecting overlapping speech, identifying speech content, or making binary/multiple-choice decisions, but the answers they produce are inconsistent with the actual audio content. Unlike hallucinated affirmatives, where models fabricate content in the absence of evidence, responses in this category remain task-relevant yet are factually incorrect. 

These errors often occur in tasks with binary or limited-choice outputs, where subtle audio cues or temporal patterns must be interpreted correctly. Common causes include overgeneralization, misalignment between model predictions and the audio timeline, or reliance on superficial heuristics instead of the underlying auditory evidence. As illustrated in Figure~\ref{TemporalComparison}, some models confidently produce answers that contradict the ground truth, while others hedge, partially refusing to commit, leading to a mix of incorrect, incomplete, or ambiguous task-grounded outputs. 

Overall, this phenomenon highlights that even when models understand the task instruction, achieving factual accuracy in audio-based reasoning remains a significant challenge.

\subsection{False Refusal Responses}

\begin{figure*}[t]
\centering
\begin{hallucinationcase}[Case 3: Speech -- Recognition]
\textbf{Prompt:}  
What does the female say in the recording?

\vspace{0.3em}
\textbf{Reference Answer:}  
No female is speaking in the recording.

\vspace{0.3em}
\textbf{Erroneous Model Outputs:}
\begin{itemize}
  \item \textbf{Qwen2-Audio-7B:} I'm sorry, I can't answer this question.
  \item \textbf{LLaMA-3.1-8B-Omni:} I'm happy to help! However, I didn't quite catch what you're referring to. Could you please provide more context or clarify your question?
  \item \textbf{LLaMA-Omni2-7B:} I'm not aware of any specific recording that matches your description. Could you provide more context or information about the recording you're referring to?
  \item \textbf{Phi-4-Multimodal:} I'm sorry, I cannot provide an answer as the audio clip is not provided.
  \item \textbf{GPT-4o-Audio-Preview:} I'm sorry, but I can't help with identifying specific speakers from audio recordings. If you need help with something else, feel free to ask!
  \item \textbf{Gemini-2.5-Flash:} I'm sorry, but I need the audio recording to tell you what the female says. Please provide the audio.
\end{itemize}

\vspace{0.3em}
\textit{This case exemplifies a \textbf{false refusal response}, where the question is answerable given the audio input, yet the models incorrectly refuse to provide an answer. Although the correct response requires recognizing the absence of a female speaker, several models default to generic uncertainty or access-related disclaimers, reflecting overly conservative refusal behavior rather than genuine ambiguity in the audio.}
\end{hallucinationcase}
\end{figure*}

This category covers cases where models refuse to answer questions that are, in fact, answerable given the provided audio input. Instead of analyzing the audio content and producing a valid response, models default to refusal behaviors, often expressed through generic uncertainty statements, access-related disclaimers, or requests for additional context. 

Such false refusals typically reflect overly conservative assumptions about model capabilities or safety constraints, rather than genuine ambiguity in the audio signal. Notably, these errors frequently arise in tasks that require recognizing the absence of a target event or attribute, where the correct answer is negative but still well-defined. As a result, models avoid committing to an answer despite sufficient evidence being available in the audio.

\subsection{Cross-domain Failure Patterns}

Beyond individual examples, we observe clear domain-dependent trends in model failure behaviors. 
For speech-related tasks, models frequently produce hallucinated affirmatives or incorrect task-grounded answers, particularly in questions involving word order, counting, or temporal relations. These errors suggest that models often rely on coarse linguistic priors or overgeneralized patterns rather than precise alignment with the underlying audio timeline.

In contrast, environmental sound tasks tend to elicit false refusals or overly cautious responses. Even when the acoustic evidence is clear, models may default to generic uncertainty or access-related disclaimers, indicating conservative assumptions about input availability or ambiguity in non-linguistic audio signals.

Music-related tasks exhibit a distinct failure profile, characterized by label confusion, task deviation, and conservative abstention. Models frequently struggle to distinguish structurally similar instruments or sound events, leading them to shift attention to irrelevant acoustic attributes or to avoid making definitive judgments. This behavior highlights persistent challenges in fine-grained acoustic reasoning and semantic grounding within complex musical contexts.

\begin{figure*}[t]
\centering
\begin{hallucinationcase}[Cross-domain Contrast: Word Order]
\textbf{Model: Qwen2-Audio-7B}

\vspace{0.3em}
\begin{itemize}
\item \textbf{Speech}  

\textit{Prompt:} Does the word "and" appear before the word "used" in the recording?

\textit{Reference:} No.

\textit{Model Output:} Yes, the word "and" appears before the word "used" in the recording. 

\textit{Issue:} Incorrect task-grounded affirmative, where the model confidently answers the word-order question but contradicts the actual temporal structure of the audio.

\vspace{0.4em}
\item \textbf{Environmental Sound}  

\textit{Prompt:} Does "Gasp" appear before "Clapping" in the recording? 

\textit{Reference:} Yes.

\textit{Model Output:} No, the "Gasp" sound appears after the "Clapping" sound in the recording.

\textit{Issue:} Incorrect task-grounded answer in temporal ordering, indicating a failure to correctly infer event sequence despite clear acoustic evidence.

\vspace{0.4em}
\item \textbf{Music} 

\textit{Prompt:} Does "ding" appear before "ding-dong" in the recording? 

\textit{Reference:} Yes.

\textit{Model Output:} I'm not sure which sound is louder just from the audio. You could try listening more closely or maybe compare the volume levels on your device. If you can tell me more about the sounds, like what they are or how they were played, that might help. So, what do you think?  

\textit{Issue:} Task deviation caused by label confusion, where the model shifts from temporal comparison to an unrelated acoustic attribute, resulting in an off-task and non-informative response.
\end{itemize}

\vspace{0.3em}
\textit{This example demonstrates that the same LALM exhibits distinct failure modes across domains when performing structurally similar word-order or temporal comparison tasks, ranging from incorrect affirmatives to task deviation and ambiguous responses.}

\end{hallucinationcase}
\end{figure*}

\end{document}